\documentclass[10pt]{article}

\usepackage[utf8]{inputenc}
\usepackage[T1]{fontenc}
\usepackage[english]{babel}
\usepackage[a4paper,nohead,twoside]{geometry}
\usepackage{amsmath}
\usepackage[mathscr]{euscript}
\usepackage{amsfonts}
\usepackage{amssymb}
\usepackage{amsthm}
\usepackage{mathtools}
\usepackage{graphicx}
\usepackage{color}
\usepackage{enumitem}
\usepackage{xspace}
\usepackage{a4wide}

\usepackage[bookmarks=true,
           bookmarksopen=true,hidelinks]{hyperref}

\newtheorem{theorem}{Theorem}[section] 
\newtheorem{definition}[theorem]{Definition}
\newtheorem{lemma}[theorem]{Lemma}
\newtheorem{corollary}[theorem]{Corollary}
\newtheorem{proposition}[theorem]{Proposition}

\theoremstyle{remark}
\newtheorem{remark}[theorem]{Remark}
\numberwithin{equation}{section}
 
  \usepackage[perpage]{footmisc}

\newcommand {\R} {\mathbb R}

\newcommand \del 		\partial

\newcommand{\diag}{\mathrm{diag}\xspace}
\renewcommand{\Re}{\mathrm{Re}\,\xspace}
\newcommand{\Eqref}[1]{Eq.~\eqref{#1}}
\newcommand{\Eqsref}[1]{Eqs.~\eqref{#1}}
\newcommand{\Appendixref}[1]{Appendix~\ref{#1}}  
\newcommand{\Sectionref}[1]{Section~\ref{#1}}  
  
\newcommand{\Defref}[1]{Definition~\ref{#1}}

\newcommand{\Lemref}[1]{Lemma~\ref{#1}}
\newcommand{\Propref}[1]{Proposition~\ref{#1}}
\newcommand{\Theoremref}[1]{Theorem~\ref{#1}}

\newcommand{\Corref}[1]{Corollary~\ref{#1}}

\newcommand{\Conditionref}[1]{Condition~\ref{#1}}
\newcommand{\Conditionsref}[1]{Conditions~\ref{#1}}
\newcommand{\keyword}[1]{\textit{#1}}

\newcommand{\zeromatrix}{\mathbf{0}}

\newcommand{\LPDE}[2]{\ensuremath{\widehat L(#1)[#2]}\xspace}
\newcommand{\LPDEu}[2]{\LPDE{u_*+#1}{#2}}

\newcommand{\Sts}[1]{\ensuremath{S^j(#1)}\xspace}

\newcommand{\St}[1]{\ensuremath{S^0(#1)}\xspace}

\newcommand{\Stna}{\ensuremath{S^0}\xspace}

\newcommand{\StLu}{\ensuremath{S^0_0(u_*)}\xspace}
\newcommand{\StLInv}[1]{\ensuremath{\left(S_0^0(#1)\right)^{-1}}\xspace}
\newcommand{\StH}[1]{\ensuremath{S^0_1(#1)}\xspace}
\newcommand{\StHu}[1]{\StH{u_*+#1}}

\newcommand{\Ssna}{\ensuremath{S^a}\xspace}
\newcommand{\Ss}[1]{\ensuremath{S^a(#1)}\xspace}

\newcommand{\N}[1]{\ensuremath{N(#1)}\xspace}

\newcommand{\Nna}{\ensuremath{N}\xspace}

\newcommand{\NLu}{\ensuremath{N_{0}(u_*)}\xspace}

\newcommand{\NODE}{\ensuremath{\mathcal N}\xspace}

\newcommand{\fna}{\ensuremath{f}\xspace}

\newcommand{\Fred}[2]{\ensuremath{\mathscr F}(#1)[#2]\xspace}
\newcommand{\Fredu}[1]{\ensuremath{\mathscr F}(u_*)[#1]\xspace}

\newcommand{\RR}[1]{\ensuremath{\mathcal{R}[#1]}\xspace}

\newcommand{\gsF}{\ensuremath{\mathcal F}\xspace}

\begin{document}

\title{A class of solutions to the Einstein equations 
\\
with AVTD behavior in generalized wave gauges}
\author{
Ellery Ames\footnote{Department of Mathematical Sciences, Chalmers University of Technology, SE-412 96 G\"oteborg, Sweden. Email: ellery.ames@chalmers.se} 
\hskip0.cm, 
Florian Beyer\footnote{Department of Mathematics and Statistics, University of Otago, P.O. Box 56, Dunedin 9054, New Zealand. Email: fbeyer@maths.otago.ac.nz.}, 
\\
James Isenberg\footnote{Department of Mathematics, University of Oregon, Eugene, OR 97403, USA. Email: isenberg@uoregon.edu.},
and 
Philippe G. LeFloch\footnote{ Laboratoire Jacques-Louis Lions \& Centre National de la Recherche Scientifique, Universit\'e Pierre et Marie Curie (Paris 6), 4 Place Jussieu, 75252 Paris, France. Email: contact@philippelefloch.org.}
}
\date{December 2016}
\maketitle

\begin{abstract}
We establish the existence of smooth vacuum Gowdy solutions, which are asymptotically velocity term dominated (AVTD) and have $T^3$-spatial topology, in an infinite dimensional family of generalized wave gauges. These results show that the AVTD property, which has so far been known to hold for solutions in areal coordinates only, is stable to perturbations of the coordinate systems. Our proof is  based on an analysis of the singular initial value problem for the Einstein vacuum equations in the generalized wave gauge formalism, and provides a framework which we anticipate to be useful for more general spacetimes.
\end{abstract}


 \vfill

\setcounter{secnumdepth}{1}
\setcounter{tocdepth}{1}

\tableofcontents

\

\

\newpage

\section{Introduction}
\label{Intro}

One of the most compelling issues in mathematical relativity concerns the nature of the boundaries\footnote{These are often referred to as spacetime ``singularities''; however, in view of the ambiguity of the term ``singularity'', we avoid that term here.} of spacetimes that  are evolved as solutions of  Einstein's equations from specified initial data sets. The work of Choquet-Bruhat and Geroch \cite{FouresBruhat:1952ji,ChoquetBruhat:1969cl} shows that for every initial data set which satisfies the Einstein constraint equations, there is a unique ``maximal development'', which is a globally hyperbolic spacetime solution of the full Einstein system, is consistent with the specified initial data, and contains all such spacetimes (up to diffeomorphism). The work of Penrose and Hawking \cite{Penrose:1965sl,Hawking:1969sw}  shows that many of the maximal development spacetimes are geodesically incomplete, and therefore have non-trivial boundaries. In some cases (e.g., the Schwarzschild solution) one cannot extend across the  boundary, and it is characterized by unbounded curvature (and consequently unbounded tidal forces). In  other cases (e.g., the Taub-NUT solutions) the boundary is a Cauchy horizon, and one can extend the spacetime smoothly into a region characterized by closed causal paths. The Strong Cosmic Censorship (SCC) conjecture (see the recent review in \cite{Isenberg:2015wr} for references)  claims that, in generic spacetimes, one cannot smoothly extend solutions beyond the maximal development.

While the issue of strong cosmic censorship is wide open for the general class of solutions of Einstein's equations, a model version of the conjecture has been proven for the family of Gowdy spacetimes (which we describe below in \Sectionref{Gowdy}). In the proof of this result \cite{Ringstrom:2009ji}, the verification that generic Gowdy solutions exhibit \emph{asymptotically velocity-term dominated} (AVTD) behavior plays an important role.\footnote{The role of AVTD behavior in proving model SCC theorems is especially evident in the proof of SCC for the polarized Gowdy spacetimes; see \cite{Isenberg:1990gn,Chrusciel:1999dk}. The idea of this role stems from the original work in \cite{Eardley:1972ig}.} Roughly speaking, a solution\footnote{Here, $M$ is the spacetime manifold, $g$ is the spacetime metric, and $\psi$ represents any non-gravitational fields. We presume that this solution is the maximal development of a set of initial data.}  $(M,g,\psi)$ of the Einstein equations has AVTD behavior if there exists a system of coordinates for $M$  such that, as one approaches the boundary of the spacetime, $(M,g,\psi)$
asymptotically approaches a spacetime $(M,\hat g,\hat \psi)$ which satisfies a system of equations (the VTD equations) which is the same as Einstein's equations but with most\footnote{But not necessarily all; see e.g., \cite{Isenberg:1990gn} or \cite{Ames:2012vz}.} of the terms containing  spatial derivatives dropped. This property is very useful for studying  SCC in  a family of spacetimes $\mathcal{A}$ which is generically AVTD because it is often easier to calculate asymptotic curvature behavior in solutions of the VTD equations than in solutions of the Einstein equations. Thus, to prove a model SCC theorem for $\mathcal{A}$, it is sufficient to first show that generic solutions of Einstein's equations in $\mathcal{A}$ are AVTD, and then show that generic solutions of the VTD equations corresponding to $ \mathcal{A}$ have unbounded curvature. It follows that generic solutions of the Einstein equations contained in $\mathcal{A}$ cannot be extended.

As noted above, to verify that a given spacetime is AVTD, one needs to find \emph{some} coordinate system in terms of which the asymptotic condition described above holds. For the Gowdy spacetimes, the \emph{areal coordinate} system is geometrically natural, and AVTD behavior has been verified using areal coordinates \cite{Kichenassamy:1999kg,Rendall:2000ki,Ringstrom:2009ji}. Does it follow that the vacuum Gowdy spacetimes are manifestly AVTD in terms of other coordinate systems as well? This is the question we address in this paper.

\Theoremref{th:maintheorem} below establishes the existence of a wide class of smooth AVTD Gowdy solutions to the vacuum Einstein equations in an infinite dimensional family of coordinates which contains the areal coordinates. While this result does not determine that Gowdy spacetimes are manifestly AVTD in terms of any choice of coordinates, it does show for the first time that the AVTD property of Gowdy spacetimes is {not} limited to areal coordinates, and that this property is in a sense stable to coordinate perturbations. The family of coordinates which we consider are generated by a certain class of gauge source functions using the generalized wave gauge formulation of the Einstein equations. 

Generalized wave coordinates (which we define below, in \Sectionref{WaveCoord}) are important for two reasons: First, unlike areal coordinates, which are well-defined only for families of spacetimes with two commuting Killing fields, generalized wave coordinates are defined for all spacetimes. Second, in terms of wave coordinates, the Einstein equations take a manifestly hyperbolic form. Such a form is essential for carrying out the analyses we use here to verify AVTD behavior.

The Gowdy family of solutions  is a useful laboratory for studying a variety of issues in mathematical relativity because the Einstein system of equations for the Gowdy family, while retaining the nonlinearities, the constraints, and the gauge freedom which mark the Einstein system generally, is relatively accessible to analysis. We already know (using areal coordinates) that generic $T^3$ Gowdy spacetimes are AVTD. The motivation for the present study is to learn about this property in systems of coordinates which can be used in wider families of solutions, such as $U(1)$-symmetric vacuum solutions (currently work in progress by us) or solutions with matter fields \cite{Beyer:2015fhs}, as well as solutions with no symmetries. 

Based on numerical studies \cite{Garfinkle:2004bq,Berger:2001dl,Berger:1998dc,Berger:1998hz} we do not expect the general class of solutions of the vacuum Einstein equations (with no symmetries imposed) to exhibit AVTD behavior in any choice of coordinates. However, such studies do suggest that polarized  $U(1)$-symmetric  solutions\footnote{The metrics of the $U(1)$-symmetric solutions which are labelled \emph{polarized} can be written as warped products of $(2+1)$-dimensional Lorentz-signature metrics with the circle orbits of the (spatially-acting) isometry group.} of the vacuum Einstein equations as well as general solutions of the Einstein-scalar field equations do exhibit AVTD behavior. The existence of \emph{analytic} spacetimes in these families of solutions which show AVTD behavior has been confirmed using analytic Fuchsian techniques \cite{Isenberg:2002ku, ChoquetBruhat:2004ix,Andersson:2001fa}. 
A major part of the motivation for the present study is to develop the tools needed to show that \emph{non-analytic} spacetimes (smooth, or with less regularity) in these families also show AVTD behavior. This is crucial because the use of the AVTD property for studying model versions of the strong cosmic censorship conjecture for a given family of spacetimes requires that AVTD behavior  be ascertained in non-analytic as well as analytic spacetimes in the chosen family.

The mathematical basis for our work here is our analysis in \cite{Ames:2012vz} of the \emph{Singular Initial Value Problem} (SIVP) for quasilinear symmetric-hyperbolic (non-analytic) PDE systems. The basic idea, we recall, is the following: We seek solutions $u:\Sigma^n \times \mathbb{R}^+ \rightarrow \mathbb{R}^d$ which satisfy the system of equations 
\begin{equation}
\label{GenEq}
\mathcal{E}[u, x, t]=0,
\end{equation}
where $\mathcal{E}$ is a function of $u$ and its (first) derivatives\footnote{For convenience, we presume that the system has been cast into first-order form.}, as well as a function of spacetime. In the standard Cauchy initial value problem for this system, we seek a solution of \eqref{GenEq} which agrees with a specified set of initial data $u_{[t_0]}:\Sigma^n \times {\{t_0\}} \rightarrow \mathbb{R}^d$, for some $t_0 \in\mathbb{R}^+.$ By contrast, in working with the singular initial value problem, we seek a solution of  \eqref{GenEq} for which $t=0$ marks the boundary of the maximal development of some (unspecified) initial data set, and which asymptotically agrees with a specified ``leading order term'' $u_*:\Sigma^n \times (0, \delta] \rightarrow \mathbb{R}^d$ for some $\delta>0$. 

We discuss in Section \ref{sec:SIVPNN}  sets of conditions on the PDE system \eqref{GenEq} and on the leading order term $u_*$ which guarantee that indeed there is a (unique) solution $u$ of $\mathcal{E}[u, x, t]=0$ which asymptotically approaches $u_*$ at the desired rate. These SIVP well-posedness results,\footnote{A well-posedness theorem for the standard Cauchy problem implies continuity of the map from initial data sets to local solutions, as well as local existence and uniqueness. We label a singular initial value problem well-posed so long as local existence and uniqueness hold; we are not concerned with continuity.}
stated in Theorem \ref{th:smoothexistenceN}, are adapted from \cite{Ames:2012vz}; note however that our statement of SIVP well-posedness here is somewhat simpler  than in \cite{Ames:2012vz}, partly as a consequence of our introducing the convenient notion of \emph{function operators} (see \Sectionref{sec:FunctionSpacesAndFunctionOperators} below) and partly because we focus here on the smooth category rather than working with less regular solutions. With some additional effort, one can show that the same techniques do also cover solutions with finite differentiability. In any case, a key feature of our work is that our results do not require real-analyticity. 

The SIVP approach is especially well-suited for proving that spacetimes admit AVTD behavior, since both focus on the asymptotic behavior of solutions of PDE systems. In particular, to show that a solution of \eqref{GenEq} has AVTD behavior, it is sufficient  to choose the leading order term $u_*$ so that it either satisfies a set of VTD equations corresponding to \eqref{GenEq} or asymptotically approaches a solution of such equations, and then show that the singular initial value problem for $\mathcal{E}[u, x, t]=0$ with leading order term $u_*$ is well-posed.

In previous studies of AVTD behavior in Gowdy spacetimes \cite{Rendall:2000ki,Beyer:2010foa,Beyer:2011ce}, areal coordinates have been chosen from the start, and the analysis has been carried out in terms of metric components and PDEs defined by a fixed areal coordinate basis. Here, working with generalized wave coordinates, we must proceed differently. Since generalized wave coordinates are defined \textit{dynamically}, through solutions of a wave-type system of equations, we work with a coupled system which combines the Einstein equations (in generalized wave coordinates) with these dynamic equations for the coordinates. It is for this combined system that we seek to set up a singular initial value problem, which we use to verify the existence of Gowdy solutions which are AVTD with respect to generalized wave coordinates.

Since the singular initial value problem plays a central role in our work here, we present a brief review of it in \Sectionref{sec:ReviewOfSIVP}. Included in this section is a review of the weighted function spaces we use in our analysis here, along with a well-posedness theorem for singular initial value problems of the sort which arise in this work. Next,  in \Sectionref{Gowdy}, we discuss generalized wave coordinates and their use as a tool for working with all solutions of Einstein's equations,  we introduce the Gowdy solutions and their representations in various coordinate systems, and we discuss particular versions of generalized wave coordinates which are useful in generating Gowdy solutions which are manifestly AVTD in terms of those coordinates. Also in this section we examine Kasner spacetimes, and use them  to help determine the asymptotic form  Gowdy solutions (metric and coordinates) should take, if they are to show AVTD behavior. This allows us to specify the appropriate form for the leading order term for our SIVP. 
Our main result is presented in \Sectionref{sec:mainresult} where we verify that the SIVP we have set up is indeed well-posed, and hence there are Gowdy solutions which are manifestly AVTD in terms of generalized wave coordinates. In \Sectionref{sec:coordinaterelationship} we discuss the solution space of our main theorem and thereby relate the AVTD solutions obtained in the previous section in generalized wave coordinates to AVTD solutions obtained in the more conventional areal coordinates.

\section{Review of the singular initial value problem}
\label{sec:ReviewOfSIVP}

As noted above, the idea of the singular initial value problem for a given PDE system is to find solutions to that system which have prescribed asymptotic behavior in the neighborhood of a designated ``boundary''  or ``singularity''. To be able to carefully define asymptotic convergence and state conditions for the well-posedness of the SIVP, we briefly review a class of time-weighted function spaces and a set of function operators on these spaces; details regarding these spaces and their properties appear in \cite{Ames:2012vz}. We  use these spaces and operators to define the sorts of equations for which the SIVP is well-posed, and we then state a well-posedness theorem for the singular initial value problem in a form which is most useful for the present work. In particular, the theorem stated here is restricted to the smooth (but not real-analytic) setting, and to operators which are rational function operators. Also, for simplicity we presume that the spatial sections of the spacetime manifold on which we work are topologically $T^n$.


\subsection{Function spaces and function operators} 
\label{sec:FunctionSpacesAndFunctionOperators}

To define the family of time-weighted Sobolev spaces, we choose 
$\mu: T^n \to \mathbb R^d$ to be a smooth
 function, we define the $d\times d$-matrix 
\begin{equation}
\label{eq:rmatrix}
\RR{\mu}(t,x) := \diag\left( t^{-\mu_1(x)},\ldots,t^{-\mu_d(x)} \right),
\end{equation}
and then for functions
 $w: (0, \delta] \times T^n \mapsto \R^d$ in $C^\infty((0,\delta]\times T^n)$ we specify the norm
\begin{align}
\label{eq:defnorm}
\begin{split}
||w||_{\delta, \mu, q} := & \sup_{t \in (0,\delta]} ||\RR{\mu} w ||_{H^q(T^n)} \\
= & \sup_{t \in (0,\delta]} \left( \sum_{|\alpha| = 0}^q \int_{T^n} |\partial^\alpha (\RR{\mu} w) |^2 dx \right)^{1/2}.
\end{split}
\end{align}
Here $H^q(T^n)$ denotes the usual Sobolev space of order $q$ on the $n$-torus $T^n$, 
$\alpha$ denotes a partial derivative multi-index, and the standard Lebesgue measure is used for the integration. 
Based on this norm, we define the function space $X_{\delta, \mu, q}(T^n)$ to be the completion of the set of functions $w \in C^\infty\left((0,\delta]\times T^n \right)$ for which the above norm is finite. Since the spatial sections are understood to be $T^n$, for convenience  we generally drop the $T^n$ argument, denoting these spaces as $X_{\delta, \mu, q}$. We let $B_{\delta, \mu, q,r}$ denote a closed ball of radius $r$ about $0$ in $X_{\delta, \mu, q}$.  To handle functions which are infinitely differentiable and for which we control all derivatives (the ``smooth case''), we also define the space $X_{\delta, \mu, \infty} := \cap_{q=0}^\infty X_{\delta, \mu, q}$. 
In the following, we refer to the quantity  $\mu$ as an \keyword{exponent vector}, or if, $d=1$, as \keyword{exponent scalar}. If we have two exponent vectors $\nu$ and $\mu$ of the same dimension, we write $\nu>\mu$ if each component of $\nu$ is strictly larger than the corresponding component of $\mu$ at each spatial point. If $\mu$ is an exponent vector and $\gamma$ an exponent scalar, then $\mu+\gamma$ refers to the exponent vector with components $\mu_i+\gamma$.

In working with $d \times d$-matrix-valued functions (such as $S^0$ in \Eqref{eq:pde} below), we use analogous norms and function spaces. In these cases, we consider $d$-vector-valued exponents $\xi$ as above and then define the space $X_{\delta,\xi,q}$ of $d\times d$-matrix-valued functions $S$ in the same way as above, but with $\RR{\mu}w$ in \Eqref{eq:defnorm} replaced by $\RR{\xi}\cdot S$ (where $\cdot$ denotes the matrix product). We note that this definition is a special case of that used  in \cite{Ames:2012vz,Ames:2013uh}; it is sufficient for our purposes here.

We next introduce a class of maps which we label as \emph{function operators}. Though not discussed in previous work \cite{Ames:2012vz,Ames:2013uh}, these function operators and their properties are very useful for discussing the regularity and asymptotic time behavior of the coefficients ($S^0, S^a, N$)  and source terms ($f$) appearing in the PDEs \Eqref{eq:pde} we consider here. Formally, we define a function operator to be a map $g$ from functions $w: (0,\delta]\times T^n\rightarrow\R^d$ to functions $g(w): (0,\delta]\times T^n\rightarrow\R^m$, where $d$ and $m$ are positive integers. A particularly important class  of such objects  may be constructed as follows. 
Let $\gamma$ be a specified continuous function 
\begin{equation}
\gamma:(0,\delta]\times T^n\times U\rightarrow\R^{m},\quad (t,x,u)\mapsto \gamma(t,x,u),
\end{equation} 
where $U$ is an open subset of $\R^{d}$. Associated to $\gamma$ is the function operator $g$ which maps functions $w:(0,\delta]\times T^n\rightarrow\R^{d}$ whose range is a subset of $U$, to functions $g(w)$ defined by
\begin{equation}
  \label{eq:functionoperatorsfunctions}
  g(w):(0,\delta]\times T^n\rightarrow\R^{m},\quad g(w)(t,x):=\gamma(t,x,w(t,x)).
\end{equation}

For our work here we need  precise control of the domain and range of such maps. To attain this, we require that the domain and range of the function operators we use here both be subsets of time-weighted Sobolev spaces of the sort defined above. Specifically, for fixed dimension index $n$ (referring to $T^n$), for exponent $d$-vector $\mu$ (possibly zero), for exponent $m$-vector $\nu$, and for differentiability index $q$ (possibly $\infty$) we define the indexed classes $\mathcal{G}_{[\delta; \mu, \nu, q]}$ of function operators as follows: First, for $\mu=0$ and for finite $q$, we have 

\begin{definition}[$\mathcal{G}_{[\delta; 0, \nu, q]}$]
  \label{def:functionoperatorsNN}
  Fix positive integers $n$, $d$, $m$ and $q>n/2$, and fix $\delta>0.$ For any real number $s_0>0$ or $s_0=\infty$, let
\begin{equation}
  \label{eq:Hspace}
  H_{\delta,q,s_0}:=\left\{\text{$w: (0,\delta]\times T^n\rightarrow\R^d$ in $X_{\delta,0,q}$}\,\, \Bigl| \, \, \sup_{t\in
       (0,\delta]} \|w(t)\|_{L^\infty(T^n)}\le s_0\right\}.
 \end{equation}
 Let $\nu$ be an exponent $m$-vector. For any $w\in H_{\delta,q,s_0}$,
we call a map $w\mapsto g(w)$ a \mbox{$(0,\nu,q)$-operator} (an element of $\mathcal{G}_{[\delta; 0, \nu, q]}$) provided that the following hold:
 \begin{enumerate}[label=\textit{(\roman{*})}, ref=(\roman{*})] 
   \item \label{cond:functionoperatorsNN1} There exists a constant
     $s_0>0$ ($s_0=\infty$ is allowed) such that for each $\delta'\in (0,\delta]$ and for all $w\in
     H_{\delta',q,s_0}$, the image $g(w)$ is a well-defined function $g(w):(0,\delta']\times T^n\rightarrow\R^m$ contained in $X_{\delta',\nu,q}$.
  \item \label{cond:functionoperatorsNN2} There exists a constant $C>0$ such that for each $\delta'\in
    (0,\delta]$ and for each $q'=q$ and $q'=q-1$, and for all $w,\tilde w\in
    H_{\delta',q,s_0}$, the following local Lipschitz estimate holds
\begin{equation}
  \label{eq:Lipschitzfunctionoperator}
  \|g(w)-g(\tilde w)\|_{\delta',\nu,q'}\le C \left(1+\|w\|_{\delta',0,q'}+\|\tilde w\|_{\delta',0,q'}\right)\|w-\tilde w\|_{\delta',0,q'}.
\end{equation}
  \end{enumerate}
\end{definition}

\noindent Before continuing on to define $\mathcal{G}_{[\delta; \mu, \nu, q]}$ and $\mathcal{G}_{[\delta; \mu, \nu, \infty]}$, we note the following: 

\begin{enumerate}
\item As mentioned above, in some cases (not all), a function operator may be specified by choosing a continuous function $\gamma: (0,\delta]\times T^n\times U\rightarrow\R^m$ and defining the map
 $w\mapsto g(w)$ as in \Eqref{eq:functionoperatorsfunctions}. However, in doing so we must be able to choose the constant $s_0$ in \Defref{def:functionoperatorsNN} so that the ranges of the functions $w\in H_{\delta,q,s_0}$ are contained in the open set $U$. If this can be done and if \Conditionref{cond:functionoperatorsNN1} holds for $\delta'=\delta$, then it automatically holds for every $\delta'\in (0,\delta]$. We shall often make use of this fact without further notice.
\item If $w\in B_{\delta,0,q,s_0/C_{q,n}}$ where $C_{q,n}$ is the
   Sobolev embedding constant for $H^q(T^n)$, then 
   \begin{equation}
   \sup_{t\in
       (0,\delta]} \|w(t)\|_{L^\infty(T^n)}\le C_{q,n} \sup_{t\in
       (0,\delta]} \|w(t)\|_{H^q(T^n)}\le s_0.
       \end{equation} 
     Hence, $w\in H_{\delta,q,s_0}$ and therefore $B_{\delta,0,q,s_0/C_{q,n}}\subset  H_{\delta,q,s_0}$. 
\item Elements of $\mathcal{G}_{[\delta; 0, \nu, q]}$ are
  uniformly bounded in the following sense: Let $w$ be an arbitrary
  function in $B_{\delta,0,q,\tilde s_0}$ for some sufficiently
  small $\tilde s_0>0$. It follows from the above remark that the map $w\mapsto g(w)$ is well-defined, and that
\[\|g(w)\|_{\delta,\nu,q}\le
\|g(0)\|_{\delta,\nu,q}+C\|w\|_{\delta,0,q}\le
\|g(0)\|_{\delta,\nu,q}+C \tilde s_0.\]
\end{enumerate}

\begin{definition}[$\mathcal{G}_{[\delta; \mu, \nu, q]}$ and $\mathcal{G}_{[\delta; \mu, \nu, \infty]}$]
  \label{def:functionoperators2}
  Fix positive integers $n$, $d$, $m$ and $q>n/2$, and fix $\delta>0.$  Let $\nu$ be an exponent $m$-vector and let $\mu$ be an exponent $d$-vector. We call the map $w\mapsto
  g(w)$ a \mbox{$(\mu,\nu,q)$-operator} (an element of $\mathcal{G}_{[\delta; \mu, \nu, q]}$) 
  if the map $w\mapsto g(\RR{-\mu} w)$ is
 a $(0,\nu,q)$-operator.
 We call the map $w\mapsto g(w)$ a $(\mu,\nu,\infty)$-operator (an element of  $\mathcal{G}_{[\delta; \mu, \nu, \infty]}$)  if
 there exists an integer $p > n/2$ such that 
$w\mapsto g(w)$ is a \mbox{$(\mu,\nu,q)$-operator} 
 for each $q\ge p$, 
 with a common constant $s_0$ for all $q\ge p$. 
\end{definition}

 In the ``smooth case'' $q=\infty$, notice that we do \textit{not} make any assumptions regarding the dependence on $q$ of the constant $C$ in \Conditionref{cond:functionoperatorsNN2} above.

It is useful for our analysis below to state a technical result which permits us in certain circumstances  to evaluate a given function operator on a function which is not a-priori known to be in the domain of that function operator (cf.\ \Conditionref{cond:functionoperatorsNN1} of \Defref{def:functionoperatorsNN}).
\begin{lemma}
  \label{lem:welldefinedFOP}
  Fix positive integers $n$, $d$ and $m$, and fix $\delta>0$. Let $\nu$ be an exponent $m$-vector and let $\mu$ be an exponent $d$-vector.
 Suppose that $g\in \mathcal{G}_{[\delta; \mu-\epsilon, \nu, \infty]}$
for all values of $\epsilon\in [0,\epsilon_0]$ where $\epsilon_0>0$ is some (possibly very  small) constant. If $w$ is any $d$-vector valued function in $X_{\delta,\mu,\infty}$, then $g$ is well-defined and $g(w)\in X_{\hat\delta,\nu,\infty}$ for some $\hat\delta\in (0,\delta]$.
\end{lemma}

We now wish to define certain special classes of function operators. All are constructed via the model described above in \Eqref{eq:functionoperatorsfunctions}; the special classes are defined by the form of the function $\gamma$. For fixed choices of positive integers $n, d$ and $N$, and for fixed $\delta>0$, if $\gamma: (0,\delta]\times T^n\times \R^d\rightarrow\R$ takes the following polynomial form 
\begin{equation}
  \label{eq:polynomialscalar}
  \gamma (t,x,u)=\sum_{i_1,\ldots,i_d=0}^N \gamma_{i_1,\ldots,i_d}(t,x)\,
  u_1^{i_1}\cdots u_d^{i_d}
\end{equation}
for some collection of  coefficient functions $\gamma_{i_1,\ldots,i_d}(t,x)$ in $X_{\delta,\tilde
  \nu_{i_1,\ldots,i_d},\infty}$ (here $\tilde\nu_{i_1,\ldots,i_d}$ is a set of scalar exponents) then we call the function operator $w\mapsto
\gamma(w)$  associated to this polynomial $\gamma$   a \keyword{scalar
  polynomial function operator}. If the function operator is constructed in this way, but with $\gamma$ being $d$-vector-valued or $d \times d$-matrix-valued, then the result is labeled a \keyword{vector (or matrix)
  polynomial function operator}. Equivalently, a vector or matrix polynomial function operator is one such that each component is a scalar function polynomial operator. 

Finally, taking quotients of polynomial function operators, we define the following class of function operators (which play a major role in our analysis here): 
\begin{definition}
\label{def:rationalFOPs}
Suppose that $h_0$ is a scalar-valued function in
$X_{\delta,\eta,\infty}$ (for some scalar exponent $\eta$) such that $1/h_0\in
X_{\delta,-\eta,\infty}$. Let $w\mapsto P_1(w)$ and $w\mapsto P_2(w)$ be a pair of scalar polynomial
function operators such that $w\mapsto P_2(w)$ is a 
$(\mu,\zeta,\infty)$-operator
for a scalar exponent $\zeta>0$. Then 
\begin{equation}
  \label{eq:scalarrationalFOPs}
  w\mapsto H(w):=\frac{P_1(w)}{ (1+P_2(w)) h_0}
\end{equation}
is called a \keyword{scalar rational function operator}. If
$w\mapsto F(w)$ is a $d$-vector-valued (or $d\times d$-matrix-valued) function operator such that
each component $w\mapsto F_{j}(w)$ (or $w\mapsto F_{ij}(w)$) is a scalar rational function
operator, then $w\mapsto F(w)$ is labeled a \keyword{vector (or matrix)
rational function operator}. 
\end{definition}

\begin{lemma}
  Suppose that $w\mapsto H(w)$ is a scalar
  rational function operator specified by \Eqref{eq:scalarrationalFOPs} for some choice of $P_1, P_2$ and $h_0$ as in Definition \ref{def:rationalFOPs}.
  Assume in addition that
  $w\mapsto P_1(w)$ is a 
  $(\mu,\nu,q)$-operator 
  for another scalar
  exponent $\nu$. Then $w\mapsto H(w)$ is a
  $(\mu,\nu-\eta,q)$-operator. 
\end{lemma}

This lemma can be proved with tools from the discussion in \cite{Ames:2012vz} and can be easily extended to vector and matrix rational function operators.


\subsection{Class of equations}
\label{classofeqns}
Our results in this paper rely on working with singular initial value problems for which the partial differential equation system can be cast into the following first-order form: 
\begin{equation}
\label{eq:pde}
\begin{split}
 \Stna(t,x,u(t,x)) Du(t,x) &+ \sum_{a=1}^n \Ssna(t,x,u(t,x)) t \partial_a u(t,x)  + \Nna(t,x,u(t,x)) u(t,x)  \\
&= \fna(t,x,u(t,x)).
\end{split}
\end{equation}
Here $u: (0, \delta] \times T^n \to \mathbb R^d$ is the vector-valued function for which the SIVP is to be solved, 
each of the $n+2$ maps $\Stna,\ldots, S^n$ and $N$  is a specified $d \times d$ matrix-valued function of the spacetime coordinates
$(t,x)\in (0, \delta] \times T^n$
 and of the unknown $u$ (but not  of the derivatives of $u$), while
$\fna=\fna(t,x,u)$ is a $\R^d$--valued function of
$(t,x,u)$.
All matrices $\Stna,\ldots, S^n$ are assumed to be symmetric.
We set $D := t \,  \partial_t= t \frac{\partial}{\partial t} =x^0 \frac{\partial}{\partial x^0}$, while $\partial_a := \frac{\partial}{\partial x^a}$ for\footnote{In all of what follows, indices $i,j,\ldots$ run over $0,1,\ldots,n$, while indices $a,b,\ldots$ take the values $1,\ldots,n$.} $a=1,\ldots,n$.
We note that while one could incorporate the term $\Nna(t,x,u) u$ into the source term $\fna(t,x,u)$, for stating the conditions we need to impose on the coefficients of the PDE \eqref{eq:pde} for well-posedness of the SIVP as well as other requirements, it is convenient to keep these terms separate. We also note that this is the form we have used in our previous studies \cite{Ames:2012vz} (for $n=1$) and \cite{Ames:2013uh} (for general $n$). For convenience, we define the differential operator $\widehat L$ as follows:
\begin{equation}
\label{eq:defLPDE}
\LPDE{u}{v}:= \sum_{j=0}^n \Sts{u} \, t \partial_j v + \N{u} v.
\end{equation}
Thus the PDE \eqref{eq:pde} can be written as 
\begin{equation}
\LPDE{u}{u}=f(u),
\end{equation}
where $f(u)$ denotes the right-hand side of \eqref{eq:pde}. 

If, for a class of initial data sets, $S^0$ is a positive-definite matrix (in the sense of eigenvalues) at each spacetime point $(t,x)$ with $t\neq0$, then the system \Eqref{eq:pde} is symmetric hyperbolic, and the corresponding Cauchy problem for initial data chosen at $t_0>0$ is well-posed. To study the singular initial value problem for \Eqref{eq:pde},
we prescribe a leading order term $u_*$ and seek a solution 
 $u=u_*+w$ for \Eqref{eq:pde} with $w$ in some specified function space with prescribed $t\searrow 0$ behavior of $w$.
Substituting $u=u_*+w$ into \Eqref{eq:pde}, one obtains a PDE system for $w$ which takes the form 
\begin{equation}
\label{eq:reducedeq}
\LPDEu{w}{w} = \Fredu{w}:= f(u_*+w) - \LPDEu{w}{u_*}.
\end{equation}
The operator $\Fredu{w}$ is often referred to as the \keyword{reduced source term operator}.
For a fixed $u_*$, the problem of existence and uniqueness for the singular initial value problem is now equivalent to establishing the 
existence and uniqueness of a solution $w$ to \Eqref{eq:reducedeq} 
in the specified function space. The key definition for studying this issue is the following:

\begin{definition}
\label{def:quasilinearlimitN}
The PDE system \Eqref{eq:pde} is  a \keyword{quasilinear symmetric hyperbolic Fuchsian system} around a specified leading order term 
$u_*\in C^\infty((0,\delta]\times T^n)$
for parameters 
$\delta>0$ and $\mu$  
if there exists a positive-definite and symmetric matrix-valued function $\StLu\in C^{\infty}(T^n)$ and a matrix-valued function $\NLu \in C^{\infty}(T^n)$, 
such that all of the following function operators are $(\mu,\zeta,\infty)$-operators for some $\zeta>0$:
\begin{align}
  \label{Ncond}
  w\mapsto N(u_*+w)-\NLu,\\
  \label{S01}
  w\mapsto \StHu{w} := \St{u_* + w} - \StLu,\\
  \label{Sa}
  w\mapsto t\Ss{u_*+w},\\
  w\mapsto \RR{\mu}\Fredu{w}.
\end{align}

If all of the function operators are rational in the sense of Definition \ref{def:rationalFOPs}, then the PDE system is labeled a \keyword{quasilinear symmetric hyperbolic rational-Fuchsian system}. If the functions $S^0(t,x,u)$, $S^a(t,x,u)$, $N(t,x,u)$ and $f(t,x,u)$ appearing \eqref{eq:pde} are all explicitly smooth, then the system is labeled a smooth quasilinear symmetric hyperbolic Fuchsian system. 
\end{definition}

While Definition \ref{def:quasilinearlimitN} appears to be different from the one given in \cite{Ames:2012vz,Ames:2013uh}, it is essentially the same. The definition given here does not involve the splitting of  $t\Ss{u_*+w}$ that is carried out in Definition 2.2 of  \cite{Ames:2012vz}; that splitting, however, is not really needed to state the (equivalent) conditions imposed on $S^0$, $S^a$, $N$, $f$ and $u_*$ in order to define a quasilinear symmetric hyperbolic Fuchsian system. We do add qualifications here -- smoothness and rationality of the function operators. However, in our work below, these qualifications hold \emph{only} if stated explicitly. We note that most of the results we prove here can be extended to finitely differential operators and to function operators which are not rational; to simplify the discussion, we impose these restrictions in our applications below.

We notice that, as a consequence of the requirement in this definition that the function operators defined in  \Eqsref{Ncond} and \eqref{S01} be $(\mu,\zeta,\infty)$-operators, it follows that
for each choice of the remainder $w$ in the specified space, the $(t,x)$-dependent functions $S^0(t,x,u_*(t,x)+w(t,x))$ and $N(t,x,u_*(t,x)+w(t,x))$ are $O(1)$ in the limit $t\searrow 0$. This is a relatively strong restriction. Indeed in practice,
to satisfy this condition it may be necessary to multiply  the whole system of equations by some power of $t$. 
Moreover, there are some example cases in which
this condition can only be satisfied if one multiplies the whole system by a \textit{matrix} of time weights; as a consequence, the symmetry of the coefficient matrices may be destroyed. Such examples suggest that  our definition of quasilinear symmetric hyperbolic Fuchsian systems may be too restrictive for some purposes. However for the application discussed in this paper, Definition \ref{def:quasilinearlimitN} is sufficient.

\subsection{Well-posedness of the singular initial value problem for Fuchsian systems}
\label{sec:SIVPNN}

The main existence and uniqueness result for the SIVP for Fuchsian systems relies
on additional structural conditions on the matrix functions appearing in \Eqref{eq:pde}. To state these conditions, we use the following definition: 
\begin{definition}[Block diagonality with respect to $\mu$]
  \label{def:BDmatrix}
 Suppose that $M: (0,\delta]\times T^n\rightarrow\R^{d\times d}$ is any smooth $d\times d$-matrix-valued function, and that $\mu$ is some $d$-vector-valued exponent.
 $M$ is called \keyword{block diagonal with respect to $\mu$} provided that 
  \[M(t,x)\RR{\mu}(t,x) - \RR{\mu}(t,x)M(t,x) = 0,\]
 (recall the definition of $\RR{\mu}$ given in \Eqref{eq:rmatrix}) for all $(t,x)\in(0,\delta]\times T^n$. 
\end{definition}
The following simple algebraic result motivates this terminology.

\begin{lemma}
  \label{lem:orderedmu}
  Let $\mu$ be a $d$-vector-valued exponent which is
  \keyword{ordered}, in the sense that 
  \begin{equation}
    \label{eq:orderedmu}
    \mu(x)=\Bigl(\underbrace{\mu^{(1)}(x),\ldots,\mu^{(1)}(x)}_{\text{$d_1$-times}},
    \underbrace{\mu^{(2)}(x),\ldots,\mu^{(2)}(x)}_{\text{$d_2$-times}},
    \ldots,
    \underbrace{\mu^{(l)}(x),\ldots,\mu^{(l)}(x)}_{\text{$d_l$-times}}\Bigr),
  \end{equation}
    where
    \begin{itemize}
    \item $l\in\{1,\ldots,d\}$,
    \item $\mu^{(i)}\not=\mu^{(j)}$ for all $i\not=j\in\{1,\ldots,l\}$,
    \item $d_1,\ldots,d_l$ are positive integers with $d_1+d_2+\ldots+d_l=d$.
    \end{itemize}
    Then any smooth $d\times d$-matrix-valued function $M$ is block
    diagonal with respect to $\mu$ if and only if $M$ is of the form
    \begin{equation}
      \label{eq:blockdiagonalmatrix}
      M(t,x)=\diag\Bigl(M^{(1)}(t,x),\ldots,M^{(l)}(t,x)\Bigr),
    \end{equation}
    where each $M^{(i)}(t,x)$ is a smooth $d_i\times
    d_i$-matrix-valued function. Moreover, if $\nu$ is any other
    $d$-vector-valued exponent with the \keyword{same ordering as $\mu$}, in the sense that 
    \[\nu(x)=\Bigl(\underbrace{\nu^{(1)}(x),\ldots,\nu^{(1)}(x)}_{\text{$d_1$-times}},
  \underbrace{\nu^{(2)}(x),\ldots,\nu^{(2)}(x)}_{\text{$d_2$-times}}
    \ldots,
    \underbrace{\nu^{(l)}(x),\ldots,\nu^{(l)}(x)}_{\text{$d_l$-times}}\Bigr),\]
    for the same integers $d_1,\ldots,d_l$, then $M$ is also block
    diagonal with respect to $\nu$.
\end{lemma}

We now use this notion of block diagonality to characterize the SIVP for \Eqref{eq:pde} with a specified leading order term $u_*$.

\begin{definition}
  \label{def:bdsystem}
 Fixing a finite integer $q>n/2+2$ and a constant $\delta>0$,
  suppose that $u_*$ is a given leading order term and $\mu$ is an
  exponent vector. The system (\ref{eq:pde}) is called  \keyword{block
    diagonal with respect to $\mu$} if, for all $u = u_*+w$ with $w\in
  X_{\delta,\mu,q}$ for which $\Sts{u}$ and 
$\N{u}$ are defined, these matrix-valued functions are
 block diagonal with respect to $\mu$.
\end{definition}

This diagonality condition is essential for deriving the energy estimates which are needed for the proof of the SIVP well-posedness theorem below. 
It ensures that both of the matrices $\Sts{u}$ and $\RR{\mu}\Sts{u}\RR{-\mu}$ are symmetric. Moreover, it guarantees that the differential  operator $\LPDE{u}{u}$ (see \Eqref{eq:defLPDE})
only couples those components of the unknown function $u$ which decay in $t$ at the same rate. 

To proceed, we assume that the system \Eqref{eq:pde} is  block
    diagonal with respect to $\mu$ (see \Defref{def:bdsystem}) and
  that $\mu$ is ordered (as in \Eqref{eq:orderedmu})
  and hence, according to
  \Lemref{lem:orderedmu}, all matrices in $\LPDE{u}{u}$
 have the same block diagonal structure as in \Eqref{eq:blockdiagonalmatrix}. In particular, the matrix
  \begin{equation}
    \label{eq:defNODE}
    \NODE=\NODE(u_*):=\StLInv{u_*}\NLu,
  \end{equation}
  is block diagonal with respect to $\mu$ in the sense of 
  \Defref{def:BDmatrix} because it has the same block 
  structure as do all
  matrices in $\LPDE{u}{u}$. Here we note that since Definition~\ref{def:quasilinearlimitN} requires that
$\StLu$ be invertible, it follows that $\NODE$ is
well-defined.
  We use 
  \begin{equation}
    \label{eq:listoflambdas}
    \Lambda:=(\lambda_1,\ldots,\lambda_d),
  \end{equation}
 to denote the list of (in general complex-valued) eigenvalues of \NODE, sorted so that the eigenvalues corresponding to each block of \NODE are listed sequentially. 

With these prerequisites established, we state a well-posedness theorem for the singular initial value problem for PDE systems of the type we consider in this work.

\begin{theorem}
\label{th:smoothexistenceN}

Suppose, for some choice of an ordered exponent vector $\mu$, a positive real number $\delta$, and a leading order term $u_*$, that \Eqref{eq:pde} is a {smooth quasilinear symmetric hyperbolic rational-Fuchsian system} around $u_*$, as specified in
\Defref{def:quasilinearlimitN}. Suppose that \Eqref{eq:pde}  is block
diagonal with respect to $\mu$ and that
\begin{equation}
  \label{eq:eigenvaluecondition}
  \mu>-\Re\Lambda,
\end{equation}
where $\Lambda$ is defined in \Eqref{eq:listoflambdas}.
Then there exists a unique solution $u$  to \Eqref{eq:pde} with remainder $w:=u-u_*$ belonging to $X_{\widetilde\delta, \mu, \infty}$ for some $\widetilde \delta \in (0, \delta]$. Moreover, $Dw \in X_{\widetilde \delta, \mu, \infty}$.
\end{theorem}

The proof of this theorem essentially follows that of Theorem~2.21 in \cite{Ames:2012vz}. As noted above, the statement of Theorem \ref{th:smoothexistenceN} here is considerably simpler than that of Theorem~2.21 in \cite{Ames:2012vz}, because the requirement here that the PDE system \Eqref{eq:pde} be \textit{rational} automatically implies the extra technical conditions which appear in the latter case.


\section{\texorpdfstring{$T^3$}{T3}-Gowdy spacetimes and generalized wave coordinates}
\label{Gowdy}

In this section, we begin by describing what generalized wave coordinates are, and how they are used in studying general solutions of Einstein's equations. 
We next  introduce the $T^3$-Gowdy spacetimes, writing them both in areal coordinates and in a general form more suited for studies involving other gauge choices. We then apply generalized wave coordinates to the Gowdy spacetimes. In doing this, we use the Kasner solutions (a subset of the $T^3$-Gowdy spacetimes) to aid us in choosing generators of generalized wave coordinates which lead to explicit AVTD behavior.

\subsection{Generalized wave coordinate gauges}
\label{WaveCoord}
The idea of the generalized wave coordinate gauges is to cast the vacuum Einstein equations $G_{ij} =0$ into an explicit  (coordinate-dependent) form which is manifestly a (quasilinear) hyperbolic PDE system for the spacetime metric\footnote{In this section, we use mid-alphabet latin letters as spacetime indices.}  $g_{ij}$. The fact that such coordinates can be chosen for any globally hyperbolic spacetime satisfying Einstein's equations depends upon the following readily-verified key observations:

I) Let $\mathcal {F}_{i}$ be any specified set of four smooth spacetime functions, let ${C_{ij}}^{k}$ (satisfying the condition ${C_{[ij]}}^{k}=0$) be any chosen set of twenty-four smooth spacetime functions,
and let  $\Gamma_{km i} := \frac 12 \left( \partial_{k} g_{m i} + \partial_{i} g_{m k} - \partial_{m} g_{ki} \right)$ and $\Gamma_{ m} := g^{ki}\Gamma_{km i}$ denote the indicated Levi-Civita connections quantities. The vacuum Einstein equations are equivalent to the (coordinate-dependent) system 
\begin{equation}
\label{eq:wgEinstEqsN}
  \begin{split}
    - \frac{1}{2} g^{kl} \partial_k\partial_l g_{i j} &+ \nabla_{(i} \mathcal F_{j )} 
    + g^{kl} g^{m n} 
    \left( \Gamma_{km i} \Gamma_{l n j} + 
      \Gamma_{km i} \Gamma_{l j n} +
      \Gamma_{km j} \Gamma_{l i n} 
    \right)\\
    &+ {C_{ij}}^{k} \mathcal (\gsF_{k}-\Gamma_{k})=0
  \end{split}
\end{equation}
if and only if $ \mathcal \gsF_{k}-\Gamma_{k} = 0.$ For any fixed choice of $\mathcal{F}_k$, \Eqref{eq:wgEinstEqsN} is a quasilinear hyperbolic system for $g_{ij}$.

II) For any set of initial data consisting of a Riemannian metric $\gamma$ and a symmetric tensor $K$ satisfying the Einstein constraint equations $G_{0j}=0$, for any spacetime metric $g$ which is compatible with this choice of initial data\footnote{$g$ is compatible with $(\gamma, K)$ in the sense that it induces  $\gamma$ as the first fundamental form and $K$ as the second fundamental form on a spacelike slice of the spacetime.}, and for any choice of the four spacetime functions $\mathcal {F}_{i}$, there exists a system of spacetime coordinates in terms of which the quantity
\begin{equation}
  \label{eq:DiDef}
  \mathcal D_{i}:= \gsF_{i}-\Gamma_{i}
\end{equation}
vanishes at $t=t_0$ (corresponding to the spacelike slice on which $g$ induces $(\gamma,K)$).

III) If, at $t=t_0$, the spacetime metric $g_{ij}$ satisfies the evolution equations \Eqref{eq:wgEinstEqsN} and induces initial data satisfying the constraints, and if coordinates have been chosen so that $\mathcal D_{i}=0$ at $t=t_0$, then it follows that $\partial_t \mathcal D_{i}=0$ at $t=t_0.$ 
This can be seen from the following relation 
\begin{equation}
\label{eq:EinsteinConstraintsAndConstraintViolationQuantities}
G^{i0}=-\frac 12 g^{00}g^{ij}\partial_t \mathcal D_{j},
\end{equation}
which is satisfied at $t = t_0$ if the metric $g_{ij}$ satisfies the Einstein evolution equations, and if coordinates have been chosen such that $\mathcal D_{i} = 0$.

IV) For any choice of the spacetime metric $g$ which satisfies \Eqref{eq:wgEinstEqsN} for given functions $\mathcal {F}_{i}$ and ${C_{ij}}^{k}$, the Bianchi identities on $g$ imply that the quantity $ \mathcal D_{i}$ satisfies the PDE system 
\begin{equation}
\label{eq:ConstraintPropagationEquationN}
 \nabla^i \nabla_i \mathcal D_{j} +{R_j}^l \mathcal D_{l}
  +\left(2\nabla_iC^{i}{}_j{}^{k} -\nabla_{j} {C_l}^{lk}\right) \mathcal D_k+
\left(2C^{i}{}_j{}^{k}- {C_l}^{lk}\delta^i_j\right) \nabla_{i} \mathcal D_k=0,
\end{equation}
where $\nabla$ is the covariant derivative compatible with the metric $g$, and $R_j{}^l$ indicates the corresponding Ricci curvature. For fixed $g$ and ${C_l}^{lk}$, this is a linear hyperbolic system for $\mathcal D_{i}$, with each of the lower-order terms containing either $\mathcal D_{k}$ or $\nabla_{i} \mathcal D_k$. Hence, for initial data  $\mathcal D_{k}(t_0,x)=0$ and $\nabla_{i} \mathcal D_k(t_0,x)=0,$ the unique solution to this system is $\mathcal D_{k}(t,x)=0$ over the whole spacetime.

With these four observations established, we may show that the Cauchy problem for the Einstein equations is well-posed as follows. We choose a smooth set of initial data $(\gamma, K)$ satisfying the constraints, and we choose the smooth spacetime functions $\mathcal {F}_{i}$ and ${C_{ij}}^{k}$. Using  observations II and III, we know that there are coordinate choices which result in initial data $(g_{ij}(t_0,x), \partial_t g_{ij}(t_0,x))$ for the system \Eqref{eq:wgEinstEqsN} having $\mathcal D_{k}(t_0,x)=0$ and $\partial_t \mathcal D_k(t_0,x)=0$. We may then treat  $(g_{ij}(t_0,x), \partial_t g_{ij}(t_0,x), \mathcal D_{k}(t_0,x)=0, \partial_t \mathcal D_{k}=0)$ as initial data for the combined hyperbolic system consisting of \Eqref{eq:wgEinstEqsN} coupled to \Eqref{eq:ConstraintPropagationEquationN}. It follows from observations I and IV that there is locally a unique solution to this initial value problem, and that the solution has $\mathcal D_{k}(t,x)=0$ over the whole spacetime. It then follows from I that the resulting spacetime metric $g_{ij}(t,x)$ is a solution of Einstein's equations.

We observe that in the above discussion the coordinate chart with coordinate functions $(t=x^0,x)$ only appears implicitly.  However it follows from the definition of the Christoffel symbols that the condition $0=\mathcal D_{k}(t,x)= \mathcal{F}_k-\Gamma_k$ can be rewritten as  $g^{ij}\partial_i\partial_j x^k =g^{kl}\mathcal{F}_l$, which is a system of wave equations for the coordinate functions $(x^k)$. Hence, the coupled hyperbolic system \eqref{eq:wgEinstEqsN}-\eqref{eq:ConstraintPropagationEquationN} may be viewed as a system of evolution equations for the metric together with the coordinates. In particular, if we express this wave equation explicitly in terms of an arbitrary local reference chart with coordinate functions $(y^k)$, it becomes an explicitly hyperbolic PDE system for the transition map $x^k(y)$:
\begin{equation}
  \label{eq:coordwaveeq}
\begin{split}
&\Box_{g_{(y)}} x^i(y)
:=g^{jk}_{(y)}(y)\left(\partial_{y^j}\partial_{y^k}x^i(y)
-(\Gamma_{(y)})_{jk}^l(y) \partial_{y^l}x^i(y)\right)\\
&
=-g^{ik}_{(x)}(x(y))\mathcal F_k (x(y)). 
\end{split}
\end{equation}
Here, $g^{jk}_{(y)}$ and $g^{ik}_{(x)}$ are the components of
the contravariant metric with respect to the $y$- and $x$-coordinates respectively. 
The functions $(\Gamma_{(y)})_{jk}^l$ are the Christoffel
symbols of the metric $g$ with respect to the $y$-coordinates.

This general setup for proving the well-posedness of Einstein's equations has been known since the work of Y. Choquet-Bruhat \cite{FouresBruhat:1952ji}. Her work uses $\mathcal {F}_i =0$, a condition which results in what has been called ``harmonic coordinates'', or equivalently ``wave coordinates''. Allowing more general choice of the functions  $\mathcal {F}_{i}$, one has ``generalized wave coordinates''.\footnote{In this paper, we use ``coordinate choice'' and ``gauge choice'' interchangeably.} Since the functions  $\mathcal {F}_{i}$ largely control the choice of coordinates, they are often labeled as the \emph{gauge source functions}.

Generalized wave coordinates are an important alternative to areal coordinates in studies of AVTD behavior via the singular initial value problem because, as seen above, for solutions which are not real analytic, it is important in working with the SIVP that the PDE system of interest be manifestly hyperbolic. 

\subsection{\texorpdfstring{$T^3$}{T3}-Gowdy spacetimes}

A $3+1$ dimensional spacetime is labeled a \emph{Gowdy spacetime} \cite{Gowdy:1974hv} if (i) it is a solution of the vacuum Einstein equations, (ii) it admits a spatially-acting $T^2$ isometry group, and (iii) the twist quantities of the Killing fields generating the isometry group vanish.\footnote{If $X$ and $Y$ are used to label the one-forms corresponding to the (commuting) Killing fields generating the isometry group, then the twist quantities vanish if and only if the four-forms  $X \wedge Y \wedge dX$ and $X\wedge Y \wedge dY$ both vanish.} The only spacetime manifolds consistent with these conditions are $\R \times T^3, \R \times S^3$ and $\R \times S^2 \times S^1,$ along with various quotients of these. We restrict our attention here to the Gowdy spacetimes on $\R \times T^3$.

The Gowdy spacetimes (especially those on $\R \times T^3$) have been used extensively to study model versions of general spacetime conjectures. It has been shown that Strong Cosmic Censorship holds for these spacetimes \cite{Ringstrom:2009ji}, that  $T^3$-Gowdy spacetimes generically exhibit AVTD behavior \cite{Ringstrom:2009ji}, that $T^3$-Gowdy spacetimes admit foliations by constant mean curvature hypersurfaces \cite{Isenberg:1982cu}, and that these spacetimes can be covered globally by areal coordinates
\cite{Moncrief:1981vo}.
Most of these studies have been carried out using the areal coordinate form of the $T^3$-Gowdy spacetime metrics, which can be written generally as follows\footnote{We note that there are other areal coordinate representations of the $T^3$-Gowdy spacetime metrics that have appeared in the literature. These are all minor variations, of little consequence.}
\begin{equation}
  \label{eq:GowdyMetric}
  g=\frac 1{\sqrt{ t}}e^{\lambda/2}(-d t^2+d x^2)
  + t\left(e^{ P} d y^2+2 e^{ P} Q d y d z + (e^{ P} Q^{2}+ e^{- P})d z^2\right),
\end{equation} 
where $\frac{\partial}{\partial y}$ and $\frac{\partial}{\partial z}$ are the Killing fields, and where $P,Q$ and $\lambda$ are functions of the coordinates $x$ and $t$.  

Without any restrictions on the coordinate gauge--except that the Killing fields be $\frac{\partial}{\partial y}$ and $\frac{\partial}{\partial z}$--the  form taken by the Gowdy metrics involves many more terms: one generally has 
\begin{equation}
\label{genGowdymetric}
g=g_{00} dt^2 + 2 g_{01}dtdx + g_{11} dx^2 + g_{AB} d\xi^A d\xi^B + g_{0A} dt d\xi^A +g_{1A} dx d\xi^A,
\end{equation}
where $\xi^2=y$ and $\xi^3 =z$, where the indices $A$ and $B$ each take the values $2$ and $3$, and where all of the metric components $g_{00}, g_{01},  g_{11}, g_{ab},  g_{0A}$ and $g_{1A}$ are functions of $t$ and $x$ (not of $y$ and $z$). For our study here, we are not concerned with showing that \textit{all} Gowdy metrics exhibit AVTD behavior in \textit{every} possible coordinate system. Hence, to simplify our analysis (without too much loss of generality) we find it useful to impose the following restrictions on the metric components:
\begin{equation}
\label{metriccond}
g_{02}\equiv g_{03}\equiv g_{12}\equiv g_{13}\equiv 0.
\end{equation}
As we see below, these conditions are preserved by the Einstein evolution equations for Gowdy metrics in the coordinate gauge choices which we introduce below in \Sectionref{wavecoordsgowdy}.
Presuming \Eqref{metriccond}, we may write
the Gowdy metric in the following form:
\begin{equation}
  \label{eq:genGowdy}
  \begin{split}
  g=&g_{00}(t,x) dt^2+2g_{01}(t,x) dtdx+g_{11}(t,x)dx^2\\
&+R(t,x)\left(E(t,x)(dy+Q(t,x)dz)^2+\frac1{E(t,x)} dz^2\right).
\end{split}
\end{equation}
We note that this form is consistent with areal coordinates --- if one chooses $R(t,x) = t, E=e^P, g_{00}= -g_{11} = -\frac{1}{\sqrt{t}}e^{\lambda/2}$ and $g_{01}=0$, then this is the areal coordinate form of the Gowdy metric --- it is, however, more general.

\subsection{Generalized wave coordinate choices for \texorpdfstring{$T^3$}{T3}-Gowdy spacetimes}
\label{wavecoordsgowdy}

While any specification of the gauge source functions $\mathcal {F}_{i}$ produces solutions in generalized wave coordinates, we focus here on certain choices which are manifestly compatible with the goal of finding Gowdy solutions which show AVTD behavior.  To determine these choices, being mindful of the central role of Kasner solutions in AVTD behavior, it is useful to recall the explicit form of the Kasner spacetimes.

The family of Kasner spacetimes consists of the set of all globally hyperbolic solutions of the vacuum Einstein equations which are spatially homogeneous with isometry group $T^3$ (also known as ``Bianchi Type I''), and generally non-isotropic. The members of the Kasner family are characterized by a single parameter $k \in \R$ (known as the \emph{asymptotic velocity}), in terms of which the Kasner metrics can be written explicitly\footnote{This is not the standard form used for the Kasner spacetimes; one usually sees them written in the form $g = -d\tau^2 + \tau^{2p_1} dx^2  + \tau^{2p_2} dy^2  + \tau^{2p_3} dz^2$, with the constraints $p_1+p_2+p_3=1$ and $(p_1)^2 + (p_2)^2 +(p_3)^2 =1$. One passes from the expression \eqref{Kasner-k} to this form using the coordinate transformation $\tau=\frac{4}{k^2+3} t^{\frac{k^2+3}4}$ and the parameter transformation $  p_1 =(k^2-1)/(k^2+3), \quad  p_2 =2(1-k)/(k^2+3),\quad p_3 =2(1+k)/(k^2+3)$.}
 in the form (on $M^{1+3}= \R^+ \times T^3$) 
\begin{equation}
  \label{Kasner-k}
  g = t^{\frac{k^2-1}{2}} \left( - dt^2 + dx^2 \right) + t^{1-k} dy^2 + t^{1+k} dz^2. 
\end{equation}
We note that for all choices of the asymptotic velocity except for $k=\pm 1$ (the flat Kasners), these spacetimes are singular (with unbounded curvature) at $t=0$. 
We also note that these coordinates are areal. Finally, we note that the Kasner spacetimes are a sub-family of the Gowdy spacetimes, characterized by the presence of an extra Killing field $\partial_x$. In particular, the Kasner spacetimes can be written in the form \Eqref{eq:genGowdy}, for some choice of the functions $g_{00}, g_{01}, g_{11}, R, E$, and $Q$.

It is straightforward to calculate the Christoffel quantities $\Gamma_i$ for the Kasner spacetimes \Eqref{Kasner-k}; one obtains 
\begin{equation}
  \label{eq:ourwavegauge}
  \Gamma_{0}=-1/t,\quad \Gamma_{1}=\Gamma_{2}=\Gamma_{3}=0.
\end{equation}
These results are the same for \emph{all} Kasner spacetimes, with no dependence on the parameter $k$. 
Recalling that a spacetime exhibits AVTD behavior in terms of a given coordinate system if the geometry seen locally by each observer asymptotically approaches that of a Kasner spacetime, the expressions for the contracted Christoffel quantities in \Eqref{eq:ourwavegauge} motivate our choice of the leading order terms for the gauge source functions $\mathcal{F}_j$ of generalized wave coordinates for the Gowdy spacetimes. 
In view  of the coupling of the gauge source functions $\mathcal{F}_j$ to the metric fields (which follows from the definition  \Eqref{eq:DiDef} of $\mathcal D_{i}$, together with the requirement that these quantities vanish), we are led to choose 
\begin{equation}
  \label{eq:asympwavegauge}
  \begin{split}
    \mathcal F_{0}(t,x,g) &= - \frac 1t + F_{0}(t,x,g), \quad \mathcal F_{1}(t,x,g) =  F_{10}(x) +F_{1}(t,x,g),\\
    \mathcal F_{2}(t,x,g) &= \mathcal F_{3}(t,x,g) = 0,
  \end{split}
\end{equation}
where $F_{0}$ is $O(t^{-1+\xi_{0}})$ and $F_{1}$ is $O(t^{\xi_{1}})$ near  $t=0$ for $\xi_{0},\xi_{1}>0$ (we provide more  precise conditions for $F_0$ and $F_1$ below) and where $F_{10}$ is a smooth function (independent of $t$). This function, which vanishes for the Kasner spacetimes, must satisfy a constraint \Eqref{eq:gaugeconstraint} involving the asymptotic metric fields. 

It is important to note that the evolution of the metric which corresponds to the choice of gauge functions 
of the form \Eqref{eq:asympwavegauge}, together with a suitable choice of the functions ${C_{ij}}^{k}$ (cf. \Eqref{eq:wgEinstEqsN}), preserves the conditions \Eqref{metriccond} along with the metric form \Eqref{eq:genGowdy} for Gowdy-symmetric metrics. As well, we readily verify that  $\Gamma_2\equiv \Gamma_3 \equiv 0$ holds for any metric \Eqref{eq:genGowdy}

The choice of gauge source functions \Eqref{eq:asympwavegauge} is \emph{not} the most general choice that  could be made for studying Gowdy solutions with AVTD behavior manifest in wave coordinates. One could, in particular, allow $\mathcal F_2$ and $\mathcal F_3$ to be non-zero, so long as they decay sufficiently quickly. We are not concerned, however, with full generality, and the choice \Eqref{eq:asympwavegauge} does simplify the analysis. Among other features, it  helps to locate the singularity at $t=0$. 

Why not simplify further,  and include the requirement that  $g_{01}=0$ among the restrictions \Eqref{metriccond} 
imposed on the metric? If this were to done, then to preserve this restriction we would need to set $F_0, F_1$, and $F_{01}$ in \Eqref{eq:asympwavegauge} to zero, hence drastically reducing the range of gauge choices. To avoid this,  we allow $g_{01}$ to be nonzero. It follows that a key part of verifying AVTD behavior in the solutions we consider here is to show that $g_{01}$ decays sufficiently rapidly.

Besides motivating the choice of gauge source functions for our analysis here, 
the Kasner metric functions also motivate our choice of the leading order terms for the metric fields $g_{00}, g_{01}, g_{11}, R, E$, and $Q$. The choice we make for these leading order terms, which encapsulate the asymptotic behavior of the metric coefficients, is spelled out in the hypothesis of \Theoremref{th:maintheorem} below\footnote{This same Kasner-motivated choice of leading order terms occurs in the areal coordinate representation of the Gowdy spacetimes, where (for example in \cite{Ringstrom:2006gy}) the leading order terms are chosen to be $\lambda=-k^2\log t$, $P=-k\log t$, $Q=0$. As well, this matching is done in studying AVTD behavior in the (half)-polarized $U(1)$-symmetric spacetimes. In the representation of \cite{Isenberg:2002ku}, after the coordinate transformation $t=e^{-\tau}$ is carried out, one has  $\phi=\frac{1+k}2\log t$, $\Lambda=\frac{k^2+2k-3}4\log t$, $x=\frac{t^{(k^2+2k-3)/2}-1}{t^{(k^2+2k-3)/2}+1}$, $\beta_a=0$ and $z$ such that $e^{4z}=1-x^2$. It is straightforward to show that  \Eqref{eq:ourwavegauge} is asymptotically satisfied in this case.}.
Recalling that the coordinates in the generalized wave gauge formulation are evolved by an inhomogeneous wave equation (with inhomogeneity $\mathcal F_i$), we leave the coefficients of this t-dependence as free functions in order to introduce, together with \Conditionref{cond:xi} of \Theoremref{th:maintheorem}, the largest possible family of coordinates which is consistent with these asymptotics.

To close this section we motivate the label \keyword{asymptotic wave gauge} for the coordinate gauges considered here and defined by the form \Eqref{eq:asympwavegauge}.
Suppose that $g_{ij}$ is a solution of Einstein's equations in the generalized wave gauge formalism with gauge source functions of the form \Eqref{eq:asympwavegauge}.
Consider any time function $t_h$ which satisfies the wave equation 
\begin{equation}
\label{eq:DefWaveTime}
\Box_g t_h=0
\end{equation}
with respect to $g$. Such a time function is called a \keyword{wave time function} (or harmonic time function), in accord with the wave gauge ($\mathcal F_i \equiv 0$) discussed above in \Sectionref{WaveCoord}. By looking for solutions $t_h$ depending only on the time function $t$ generated by the generalized wave gauge source functions \Eqref{eq:asympwavegauge}, and presuming that the shift decays sufficiently fast, we determine that the solutions to the ordinary differential equation implied by \Eqref{eq:DefWaveTime} show that $t_h$ is related logarithmically to $t$. 
Since this holds for \textit{any} gauge source functions \Eqref{eq:asympwavegauge} at least asymptotically, (and presuming that the shift variable decays sufficiently fast close to $t=0$), we call any set of gauge source functions \Eqref{eq:asympwavegauge} \keyword{asymptotic wave gauge source functions}.


\section{AVTD \texorpdfstring{$T^3$}{T3}-Gowdy vacuum solutions in asymptotic wave gauges}
\label{sec:mainresult}

The main result of this paper is that there is a fairly wide class of $T^3$-Gowdy spacetimes which exhibit AVTD behavior in a fairly wide class of generalized wave coordinates. While not precluding the possibility that such behavior is found in  even larger classes of such spacetimes and such coordinates, \Theoremref{th:maintheorem} (our main result) carefully states what we mean by these ``fairly wide classes'' in terms of the free choice of certain functions and certain numbers which parametrize the asymptotic data for these spacetimes and for the gauge source functions. We present the detailed statement of \Theoremref{th:maintheorem} in Subsection \ref{mainthm} of this paper, along with clarifying comments. In Subsection \ref{sec:outlinemainproof}, we outline the main steps of the proof of \Theoremref{th:maintheorem}. Then in Subsection~\ref{constructionof}, we carry out the portion of the proof which uses a singular initial value problem formulation to construct  these spacetimes and their coordinates, in Subsection~\ref{SptmsAreSolns}, we show that these spacetimes are solutions of the vacuum Einstein equations, and in Subsection~\ref{sec:AVTDGowdy} we complete the proof by verifying that these solutions do indeed exhibit AVTD behavior. Certain of the technical calculations needed for the proof are included in the Appendices.

\subsection{Main result}
\label{mainthm}

We state our main result, Theorem \ref{th:maintheorem}, by first listing the choices of parametrizing functions--which we collectively label $\mathcal{P}$--one
makes to specify a particular Gowdy solution which is AVTD in  terms of a particular set of wave coordinates. These parametrizing functions are all defined as smooth functions either on the circle (with coordinate $x$), or on an interval cross the circle (with coordinates $(t,x)$). We note here a change in notation from that used above in our review of the Singular Initial Value Problem. In that  review, in Section \ref{sec:ReviewOfSIVP}, the exponent vector for a remainder function $w$  is denoted by $\mu$.  Here in Section \ref{sec:mainresult}, it is useful to instead denote this exponent vector by $\kappa + \mu$, where $\kappa$ is the exponent vector for the leading order term, and where $\mu > 0$.

\begin{theorem}[Existence of AVTD Gowdy vacuum solutions in asymptotic wave gauges]
  \label{th:maintheorem}
Let the space $\mathcal{P}$ consist of the following functions:
  \begin{enumerate}[label=\textit{(\roman{*})}, ref=(\roman{*})] 
  \item \label{cond:k} 
    \keyword{Asymptotic velocity}:
    A function $k\in C^\infty(T^1)$ such that $0<k(x)<3/4$ for all $x\in T^1$.
  \item \label{cond:dataTh} 
    \keyword{Asymptotic metric data}:
    A set of functions $g_{11*}$, $R_{*}, E_*, Q_*, Q_{**}\in C^\infty(T^1)$,  with
    $R_{*}, E_*, g_{11*}>0$, collectively satisfying the constraint\footnote{The origin of this constraint is explained 
      below; see \Eqref{eq:RSconst}, together with the discussion immediately following.}
 \begin{equation}
      \label{eq:integralconstraintN}
        \int_0^{2\pi}\left(
        -k(x)\frac{ E_*'(x)}{E_*(x)}+2 k(x) E_*^2(x) Q_{**}(x) Q_{*}'(x)+\frac{3-k^2(x)}{2}\frac{R_*'(x)}{R_*(x)}\right)dx=0,
    \end{equation}
    along with a positive constant  $g_{00**}>0$.
    
  \item \label{cond:xi} \keyword{Asymptotic gauge source function data}: A pair of functions $F_0 \in X_{\delta,-1+\xi_0,\infty}\cap C^\infty((0,\delta]\times T^1)$ and $F_1 \in X_{\delta,\xi_1,\infty}\cap C^\infty((0,\delta]\times T^1)$, for some $\delta>0$ and for a pair of  exponent functions $\xi_0$, $\xi_1$ with $\xi_0(x)>\max\{0,2k(x)-1\}$ and with $\xi_1(x)>0$ for all $x\in T^1$.

  \end{enumerate}
For any given choice of an element in $\mathcal{P}$ (i.e., for any choice of the functions and constants listed above), construct the functions 
\begin{equation}
    \label{eq:RSconst}
    g_{00*}(x):=-g_{00**} e^{\int_0^{x}\bigl(
        -k(\xi)\frac{ E_*'(\xi)}{E_*(\xi)}+2 k(\xi) E_*^2(\xi) Q_{**}(\xi) Q_{*}'(\xi)+\frac{3-k^2(\xi)}{2}\frac{R_*'(\xi)}{R_*(\xi)}\bigr)d\xi},
        \end{equation} 
and
\begin{equation}
   \label{eq:gaugeconstraint}
      F_{10}(x):=-\frac{g_{00*}' (x)}{2g_{00*}(x)}+\frac{g_{11*}' (x)}{2g_{11*}(x)}-\frac{R_*' (x)}{R_*(x)}, 
    \end{equation}
 for all $x\in T^1$. (The function $g_{00*}(x)$ is well-defined on $T^1$, as a consequence of \Eqref{eq:integralconstraintN}.)        
        
        Then there exists a $\hat\delta>0$, an exponent vector $\mu=(\mu_1,\ldots,\mu_6)>0$
  and a unique smooth Gowdy symmetric Lorentzian metric $g$ which satisfies Einstein's
  vacuum equations and which, for the coordinate gauge choice determined by the gauge source functions
  \begin{equation}
  \label{eq:asympwavegaugemainthm}
  \begin{split}
    \mathcal F_{0}(t,x) &= - \frac 1t + F_{0}(t,x), \quad \mathcal F_{1}(t,x) =  F_{10}(x) +F_{1}(t,x),\\
    \mathcal F_{2}(t,x) &= \mathcal F_{3}(t,x) = 0,
  \end{split}
\end{equation}
has metric components taking the following form:
  \begin{align}
    \label{eq:leadinorderterms}
    g_{00}(t,x)&=g_{00*}(x)t^{(k^2(x)-1)/2}+w_{00}(t,x),\\    
    g_{11}(t,x)&=g_{11*}(x)t^{(k^2(x)-1)/2}+w_{11}(t,x),\\
    \label{eq:leadinordertermsshift}
    g_{01}(t,x)&=w_{01}(t,x),\\
    \label{eq:blockdiagmetric}
    g_{02}&\equiv g_{03}\equiv g_{12}\equiv g_{13}\equiv 0,\\
    \label{eq:g22}
    g_{22}(t,x)&=R(t,x) E(t,x),\\
    \label{eq:g23}
    g_{23}(t,x)&=R(t,x) E(t,x)(Q_*(x)+Q(t,x)),\\
    \label{eq:g33}
    g_{33}(t,x)&=R(t,x) E(t,x) (Q_*(x) +Q(t,x))^2+R(t,x)/E(t,x),
    \end{align}
    and
    \begin{align}
      R(t,x)&=R_*(x) t+w_R(t,x),\\
      E(t,x)&=E_*(x) t^{-k(x)}+w_E(t,x),\\
      \label{eq:leadinordertermsII}
      Q(t,x)&=Q_{**}(x) t^{2k(x)}+w_Q(t,x).
    \end{align}
    The remainders satisfy the fall-off conditions
    \begin{equation}
      \label{eq:remainderstheorem}
      w_{00}\in X_{\hat\delta,(k^2-1)/2+\mu_1,\infty}, w_{11}\in
      X_{\hat\delta,(k^2-1)/2+\mu_2,\infty}, w_{01}\in
      X_{\hat\delta,(k^2+1)/2+\mu_3,\infty}, 
    \end{equation}
    and
    \begin{equation}
      w_R \in X_{\hat\delta,1+\mu_4,\infty}, w_E \in
    X_{\hat\delta,-k+\mu_5,\infty},
    w_Q \in X_{\hat\delta,2k+\mu_6,\infty}.
  \end{equation}
  The same respective spaces describe time derivatives $D^lw_{00}, D^l w_{11}, D^l w_{01}, D^l w_{R}, D^lw_{E}$, and $D^l w_Q$ of arbitrary order $l\ge 0$. 
  This metric $g$ is AVTD with respect to the coordinates generated by the gauge choice \Eqref{eq:asympwavegaugemainthm}.   
\end{theorem}

Before carrying out the proof of this theorem (in Subsections~\ref{constructionof}, \ref{SptmsAreSolns} and \ref{sec:AVTDGowdy} below), we make a few comments: 

\begin{remark}
\label{Countingfunctions}
\Theoremref{th:maintheorem} shows that for each choice of an element of $\mathcal{P}$--i.e., for each choice of the asymptotic data listed in (i)-(ii)-(iii) above--there is a vacuum solution to the Einstein equations which has Gowdy symmetry and which exhibits AVTD behavior in one of a large family of wave coordinate gauges.  
The number of free functions comprising the asymptotic parametrizing data $\mathcal{P}$ for specifying these spacetimes and their coordinate systems is larger than that needed to specify AVTD Gowdy spacetimes in areal coordinates, which are discussed in
\cite{Rendall:2000ki,Stahl:2002bv,Beyer:2010foa}. The areal coordinate case corresponds to the special case of \Theoremref{th:maintheorem} if one specifies 
$g_{00**}=1$, $R_*=1$, $F_{10}=0$ and $F_0\equiv F_1\equiv 0$, and where all other data functions are subject to the standard areal Gowdy constraint
\[\int_0^{2\pi}\left(
        -k(x)\frac{ E_*'(x)}{E_*(x)}+2 k(x) E_*^2(x) Q_{**}(x) Q_{*}'(x)\right)dx=0.\]
The two constraints \Eqsref{eq:RSconst} and \eqref{eq:gaugeconstraint} then imply that
\begin{equation*}
  -g_{00*}(x)=g_{11*}(x)= e^{\int_0^{x}\bigl(
    -k(\xi)\frac{ E_*'(\xi)}{E_*(\xi)}+2 k(\xi) E_*^2(\xi) Q_{**}(\xi) Q_{*}'(\xi)\bigr)d\xi}.
\end{equation*} 
These expressions take the usual areal coordinate form if we identify $E_*(x)=e^{P_{**}(x)}$ and $g_{00*}(x) = e^{\lambda_{**}(x)/2}$. One finds that the corresponding solution described by our theorem has the property $R\equiv t$, $g_{00}\equiv -g_{11}$ and $g_{01}\equiv 0$. 

A larger subset of the solutions obtained from \Theoremref{th:maintheorem} have coordinates which are ``asymptotically areal'' in the sense that the area function of the 2-surfaces of symmetry \textit{approaches} the time coordinate (or a constant multiple thereof). This subset of solutions is therefore determined by taking the asymptotic data function $R_*(x)$ to be unity (or to be some other positive constant).

We emphasize,  however that \Theoremref{th:maintheorem} also establishes the local existence of AVTD solutions in coordinates which are neither areal nor asymptotically areal. To the authors' knowledge this is the first result to this effect. Since Ringstr\"om \cite{Ringstrom:2009ji} has considered generic Gowdy solutions and his results have been obtained using areal coordinates, one might guess that the non-areal solutions we find here are in fact diffeomorphic to Gowdy solutions in areal coordinates. In \Sectionref{sec:coordinaterelationship} we discuss, in particular, the relationship between areal coordinates and the general class of coordinates gauges considered in our theorem. The question as to whether \emph{every} solution obtained via \Theoremref{th:maintheorem} is diffeomorphic to a solution  known to be AVTD in terms of  areal coordinates remains open, however. In any case, the most important consequence of our theorem is that the inherently coordinate-dependent notion of AVTD behavior in solutions of Einstein' equations is stable under changes of the coordinates, at least if the restrictions of \Theoremref{th:maintheorem} are imposed.
\end{remark} 
 
\begin{remark}
\label{nometricdependence}
We observe that the asymptotic data for the gauge source functions, as described in condition (iii) of the hypothesis of \Theoremref{th:maintheorem}
 depend only on the spacetime coordinates and are, in particular, independent of the metric fields. This simplification is purely for the convenience of presentation. In fact, this restriction can be relaxed so long as the gauge source functions satisfy the more general 
restriction, listed as condition (v) of the hypothesis of \Propref{prop:existenceevoleqs}. The reason  we stick with the simpler version in \Theoremref{th:maintheorem} is that it is cumbersome to express the more general condition without the ``first-order variables'' introduced in 
\Sectionref{constructionof} below.

\end{remark}

\begin{remark}
\label{kcondition}
In  studies of AVTD behavior in Gowdy spacetimes in areal coordinates, the  restriction on the asymptotic velocity
$k$ generally  imposed has been $0<k<1$. Here, in \Theoremref{th:maintheorem}, we require $0<k<3/4$. 
We believe that this  is not  a real difference, and that this restriction could be loosened. Indeed, in some of the earlier studies of Gowdy spacetimes in areal gauge \cite{Rendall:2000ki,Stahl:2002bv}, a similar restriction on $k$ is imposed. In these works, this restriction is loosened using a successively improved sequence of leading order terms. We believe that the same procedure could be applied here. However, since the analysis is significantly more complicated in the wave gauge formalism, we refrain from verifying this.

In the polarized ($Q_*=Q_{**}=0$) and half-polarized  ($Q_*=\textrm{const}$) cases, no additional arguments are necessary to make this restriction on $k$ disappear. This is so because certain problematic terms in the Einstein evolution equations are then identically zero.  In these special cases, $k$ is allowed to be an arbitrary function.
\end{remark}

\subsection{Outline of the proof of the main result}
\label{sec:outlinemainproof}

The proof of \Theoremref{th:maintheorem} consists of the following three main tasks: 1) Showing that for any choice of a set of asymptotic data in $\mathcal{P}$, the singular initial value problem corresponding to \Eqref{eq:wgEinstEqsN} for the metrics with Gowdy symmetry is well-posed. 2) Showing that for any such choice of asymptotic data, it is also true that the singular initial value problem for the constraint-violation quantities $\mathcal{D}_i$ (see \Eqref{eq:DiDef}) is well-posed, with solutions that necessarily vanish. 3) Showing that for any such  chosen asymptotic data, the Gowdy solution whose existence and wave-coordinate representation follows from the first two tasks must exhibit AVTD behavior in those coordinates. We now outline, in a bit more detail, the concrete steps that must be carried through in order to accomplish these tasks and thereby prove \Theoremref{th:maintheorem}. We label these steps in accord with these three main tasks.

\emph{Step 1a:} The starting point for the proof, is the substitution into \Eqref{eq:wgEinstEqsN}
 of the various expressions \Eqref{eq:leadinorderterms}-\Eqref{eq:leadinordertermsII} for the metric in terms of asymptotic data and remainder quantities, and \Eqref{eq:asympwavegaugemainthm} for the gauge source functions in terms of asymptotic data and remainder terms. This produces a second-order system (parametrized by asymptotic data) for the remainder terms. Since our formulation of the singular initial value problem (see Section \ref{sec:ReviewOfSIVP}) works  rather with first-order PDE systems, we proceed by introducing new functions corresponding to the first derivatives of the remainder terms, thereby producing  a (triple in size) first-order system. In doing this, we verify that the system is symmetric hyperbolic.
 
 \emph{Step 1b:} If we combine the first-order PDE system for the remainder terms (together with their first derivatives) obtained in Step 1a  with the leading order terms corresponding to the choice of asymptotic data (an element of $\mathcal{P})$, we obtain a singular initial value problem. In this step, we verify that this singular initial value problem satisfies the hypotheses of Theorem \ref{th:smoothexistenceN}, and therefore is well-posed. The statement of this verification appears in Proposition \ref{prop:existenceevoleqs} below.
 
 \emph{Step 1c:} Using the existence and uniqueness results obtained in Step 1b for the first-order PDE system,  we determine in this step that existence and uniqueness hold for solutions of the original second-order system. It follows that for the chosen set of asymptotic data (an element of $\mathcal{P}$), there exists a unique spacetime and a unique set of wave coordinates (in a neighborhood of the singularity) such that the components of the metric in terms of these coordinates satisfy \Eqref{eq:wgEinstEqsN}. These results are stated in Proposition \ref{prop:existenceevoleqs2ndorder} below.

It is not true \emph{a priori} that the spacetime whose existence and uniqueness are verified in Step 1c is a solution of the vacuum Einstein equations. To show this, it is sufficient to prove that for any choice of the asymptotic data consistent with the hypothesis of \Theoremref{th:maintheorem}, the constraint violation quantities $\mathcal{D}_i$ vanish. As noted above, such a result follows if (i) the asymptotic data for $\mathcal{D}_i$ vanish, and (ii) the singular initial value problem for $\mathcal{D}_i$ corresponding to (a first-order version of) \Eqref{eq:ConstraintPropagationEquationN} has a unique solution. 
There is a subtlety involved in doing this which we explain in detail later. It turns out that we are only able to find sufficient conditions for both (i) and (ii) if we
tighten the conditions on the metric and gauge source function asymptotic data. 
The definition of $\mathcal{P}$ in the hypothesis for \Theoremref{th:maintheorem} includes this tightening. As part of this process, the next step of the proof is a somewhat technical lemma: 

\emph{Step 2a:} Using the conditions on the asymptotic data imposed by conditions (i)-(iii) in \Theoremref{th:maintheorem}, we prove in this step that the ``shift quantity'' $g_{01}$ decays rapidly (at a rate described in Proposition \ref{prop:firstimprovement}) as $t$ approaches the singularity (marked by $t=0$). 
 
 \emph{Step 2b:} We set up a singular initial value problem for $\mathcal{D}_i$ based on a first-order version of \Eqref{eq:ConstraintPropagationEquationN}, together with asymptotic data for $\mathcal{D}_i$ and its first time derivative. We show that it follows from conditions (i)-(iii) in \Theoremref{th:maintheorem} that this asymptotic data vanishes. We then 
use the decay rate established in Step 2a to verify (based on Theorem \ref{th:smoothexistenceN}) that this singular initial value problem is well-posed. The consequent existence and uniqueness for solutions of this problem then implies that since $\mathcal{D}_i=0$ is a solution, it is necessarily the only solution. We conclude that the spacetimes whose existence is guaranteed in Step 1c must be solutions of the vacuum Einstein equations.

\emph{Step 3:} To complete the proof of \Theoremref{th:maintheorem}, we show in this step that for each choice of a set of asymptotic data (contained in the set $\mathcal{P}$), the Gowdy spacetime constructed via the singular initial value problem using this data, and represented in wave coordinates also generated from this data, must asymptotically approach a solution of a truncated ``VTD'' version of the Einstein equations. It follows that each such spacetime exhibits AVTD behavior in terms of these wave coordinates. 

\subsection{Construction of the spacetimes and coordinates}
\label{constructionof}

We begin carrying out the details of the proof in this subsection. Here, we focus on constructing the spacetimes along with the generalized wave coordinates, in terms of which the spacetime metric fields are represented. In doing this, we follow the steps of the outline presented above.  

\textbf{Carrying out Step 1a:} 
We consider the Einstein evolution equations, \Eqref{eq:wgEinstEqsN}, with the choice of gauge source functions specified in \Eqref{eq:asympwavegauge}. It is useful to set $ {C_{ij}}^k(t,x)=0$ for all $i,j,k$ except for the following 
\begin{equation}
  \label{eq:choiceconstraintmultiples}
  {C_{00}}^0(t,x)=\frac{\gamma_0(x)}t, \quad {C_{01}}^1(t,x)={C_{10}}^1(t,x)=\frac{\gamma_1(x)}t,
\end{equation}
for as yet unspecified smooth functions $\gamma_0$ and
$\gamma_1$. The resulting PDE system takes the form
\begin{equation}
  \label{eq:wavepdeNGowdy}
      \sum_{k,l=0}^1g^{kl}\partial_{x^k}\partial_{x^l}g_{ij}= 2\widehat H_{ij},  
\end{equation}
with
\begin{equation}
\label{eq:wavepdeNGowdy2}
\widehat H_{ij}:=\nabla_{(i} \mathcal F_{j )} + g^{kl} g^{m n} 
  \left( \Gamma_{km i} \Gamma_{l n j} + 
         \Gamma_{km i} \Gamma_{l j n} +
         \Gamma_{km j} \Gamma_{l i n} 
  \right)
+ {C_{ij}}^{k} \mathcal D_{k}.
\end{equation}

We first argue that the metric restrictions 
\begin{equation}
  \label{eq:6N}
  g_{02}\equiv g_{03}\equiv g_{12}\equiv g_{13}\equiv 0,
\end{equation}
(cf.~\Eqref{eq:blockdiagmetric}) are preserved by the Einstein evolution equations with $T^2$ isometry (and with $\frac{\partial}{\partial x^2}$ and $\frac{\partial}{\partial x^3}$ as Killing fields). We verify this by showing that the quantities 
$\widehat H_{ij}$ for $ij=02, 03, 12,$ and $13$ all vanish if we substitute the conditions \eqref{eq:6N} into the formula 
\eqref{eq:wavepdeNGowdy2} for $\widehat H_{ij}$. We note that conditions enforce the vanishing of the $T^2$-symmetry twist quantities, and therefore guarantee that the spacetimes under study are indeed Gowdy spacetimes. We also note that 
\eqref{eq:6N} does not further restrict our work here to a subfamily of the Gowdy spacetimes. 

We proceed to work with the Einstein equations for the remaining metric components $g_{00}$, $g_{01}$, $g_{11}$, $g_{22}$, $g_{23}$ and $g_{33}$. The latter three are parametrized as in \Eqsref{eq:g22} -- \eqref{eq:g33}. We presume that $Q_*$ is a given smooth function and hence work 
with the following vector consisting of six unknown functions: 
\begin{equation}
  \label{eq:defu}
  u(t,x)=\left(g_{00}(t,x),g_{11}(t,x),g_{01}(t,x),R(t,x), E(t,x), Q(t,x)\right)^T.
\end{equation}
It is straightforward  to show that the system of wave equations for the metric components $g_{ij}$, \Eqsref{eq:wavepdeNGowdy} -- \eqref{eq:wavepdeNGowdy2}, implies a similar system of wave equations for the components of this unknown vector $u$ of the form \Eqref{eq:wavepdeN} with $d=6$ and $n=1$; i.e.,
\begin{equation}
\label{eq:waveblubb}
      \sum_{k,l=0}^1g^{kl}\partial_{x^k}\partial_{x^l} u= 2 H,  
\end{equation}
where the vector $H$ can be computed explicitly from previous expressions. 

We now wish to convert this second-order system into a first-order symmetric hyperbolic system. This is achieved by introducing the $18$-dimensional vector $U$ as in \Eqsref{eq:firstordervariables} and \eqref{eq:firstordervariables2}; i.e., we set 
\begin{equation}
   \label{eq:firstordervariablesGowdy}
    U_{-1}^i:=u^i,\quad U_0^i:=Du^i-\alpha u^i,\quad U_1^i:=t\partial_{x}
    u^i,
    \quad U^i:=(U^i_{-1},U_0^i,U_1^i)^T,
  \end{equation}
  for $i=1,\ldots,6$,
  with $\alpha$ a constant to be fixed below, and we define 
  \begin{equation}
   \label{eq:firstordervariables2Gowdy}
    U:=(U^1,\ldots,U^6)^T.
  \end{equation}
 As discussed in \Appendixref{sec:2ndorder21storder}, the second-order system for $u$ above implies a first-order system for the extended vector $U$ of the form \Eqsref{eq:firstordersystem} --
  \eqref{eq:firstordersystem2}; i.e.,
\begin{equation} 
  \label{eq:firstordersystemGowdy}
    S^0 DU+S^1 t\partial_{x} U+\tilde N U=\tilde f[U],
   \end{equation}
  with block-diagonal matrices
  \begin{equation}
    \label{eq:block2fullGowdy}
    S^0=\diag (\mathbf s^0,\ldots,\mathbf s^0),\quad
    S^1=\diag (\mathbf s^1,\ldots,\mathbf s^1),\quad
    \tilde N=\diag(\tilde{\mathbf n},\ldots,\tilde{\mathbf n}).
  \end{equation}
  
 The general form of the blocks $\mathbf s^i$ is given in \Eqref{eq:s0}. Recall that $g^{ij}$ are the components of the inverse matrix of $(g_{ij})$. In our present application we find  
\begin{equation}
  \label{eq:s0Gowdy}
  \mathbf s^0=
    \begin{pmatrix}
      1 & 0 & 0\\
      0 & 1 & 0\\
      0 & 0 & -g^{11}/g^{00}
    \end{pmatrix}=
    \begin{pmatrix}
      1 & 0 & 0\\
      0 & 1 & 0\\
      0 & 0 & -g_{00}/g_{11}
    \end{pmatrix}
    =
    \begin{pmatrix}
      1 & 0 & 0\\
      0 & 1 & 0\\
      0 & 0 & -U^1_{-1}/U^2_{-1}
    \end{pmatrix},
  \end{equation}
 and
\begin{equation}
  \label{eq:s1Gowdy}
  \begin{split}
  \mathbf s^1
    &=
    \begin{pmatrix}
      0 & 0 & 0 \\
      0 & 2g^{01}/g^{00} & g^{11}/g^{00} \\
      0 & g^{11}/g^{00} & 0
    \end{pmatrix}
    =
    \begin{pmatrix}
      0 & 0 & 0 \\
      0 & -2g_{01}/g_{11} & g_{00}/g_{11} \\
      0 & g_{00}/g_{11} & 0
    \end{pmatrix}\\
    &=
    \begin{pmatrix}
      0 & 0 & 0 \\
      0 & -2U^3_{-1}/U^2_{-1} & U^1_{-1}/U^2_{-1} \\
      0 & U^1_{-1}/U^2_{-1} & 0
    \end{pmatrix},
  \end{split}
\end{equation}
  while
\begin{equation}
\begin{split}
\label{eq:tildenGowdy}
\tilde{\mathbf n}&=
    \begin{pmatrix}
      -\alpha & -1 & 0 \\
      -(1-\alpha)\alpha & -1+\alpha & 0 \\
      0 & 0 & (1+\alpha) g^{11}/g^{00} 
    \end{pmatrix}\\
    &=
    \begin{pmatrix}
      -\alpha & -1 & 0 \\
      -(1-\alpha)\alpha & -1+\alpha & 0\\
      0 & 0 & (1+\alpha) g_{00}/g_{11} 
    \end{pmatrix}\\
    &=
    \begin{pmatrix}
      -\alpha & -1 & 0 \\
      -(1-\alpha)\alpha & -1+\alpha & 0\\
      0 & 0 & (1+\alpha) U^1_{-1}/U^2_{-1}
    \end{pmatrix}.
  \end{split}
\end{equation}
Moreover, we have ($i=1,\ldots,6$),
  \begin{equation}
    \label{eq:firstordersystem2Gowdy}
    \tilde f[U]^i=
    \left(0, \frac{2t^2 \Xi }{U^2_{-1}} H^i+2\alpha \frac{U^3_{-1}U^i_1}{U^2_{-1}},0\right)^T,
  \end{equation}
cf.~\Eqref{eq:firstordersystem2}, where $H^i$ are the components of the $6$-dimensional vector $H$ in \Eqref{eq:waveblubb}, and where 
\[\Xi(t,x) =g_{00}g_{11}-g_{01}^2=U^1_{-1}U^2_{-1}-(U^3_{-1})^2.\]

We verify by inspection that this first-order PDE system is symmetric hyperbolic. In the remainder of the paper, we refer to this system as 
the \textit{first-order evolution system}.

\textbf{Carrying out Step 1b:}
The aim is now to construct solutions of the first-order system \Eqsref{eq:firstordersystemGowdy} -- \eqref{eq:firstordersystem2Gowdy} with the leading-order behavior asserted in \Theoremref{th:maintheorem}. To do this, it is sufficient  to show that for a choice of a leading order term $U_*$ for $U$ which is compatible with the conditions in \Theoremref{th:maintheorem}, and with a choice of the function space for the remainder term  which is also compatible with \Theoremref{th:maintheorem}, the resulting singular initial value problem satisfies the conditions of \Theoremref{th:smoothexistenceN} and is consequently well-posed.

Mindful of conditions \Eqref{eq:leadinorderterms}-\eqref{eq:leadinordertermsII}, we choose the leading order term for the components of the vector $u_*$ (see \Eqref{eq:defu}) in the form
\begin{equation}
  \label{eq:vecuSmetric} 
  \begin{split}
    u_*(t,x)=\Bigl(g_{00*}(x) t^{(k^2(x)-1)/2},g_{11*}(x) t^{(k^2(x)-1)/2},0,\\
    R_*(x) t, E_*(x)t^{-k(x)}, Q_{**}(x)t^{2k(x)}\Bigr)^T
  \end{split}
\end{equation}
for asymptotic data $g_{00*}$, $g_{11*}$, $R_*$, $E_*$ and $Q_{**}$ in $C^\infty(T^1)$ and for an asymptotic velocity $k\in C^\infty(T^1)$.  We note that at this stage, we do not require that the asymptotic data functions satisfy the conditions in the hypothesis of \Theoremref{th:maintheorem}. For the vector $U$ used in the formulation of the  first-order system (see \Eqsref{eq:firstordervariablesGowdy}-\eqref{eq:firstordervariables2Gowdy}), we choose the leading order term $U_*$ to take the form
\begin{equation}
    \label{eq:defUSGowdy}
    U_*:=(U^1_*,\ldots, U^6_*)^T,\quad
    U^i_*:=(u^i_*, Du^i_*-\alpha u^i_*,0)^T,
\end{equation}
for $i=1,\ldots,6$. 

To set up the function spaces for the remainder term, we next define the $\R^6$-vector 
\begin{equation}
\label{eq:kappak}
\kappa:=(\kappa_1,\ldots,\kappa_6)
=\left((k^2-1)/2,(k^2-1)/2,(k^2-1)/2,1,-k,2k\right),
\end{equation}
from which we construct the $\R^{18}$-vector
\begin{equation}
  \label{eq:defKappaGowdy}
  \hat\kappa:=(\kappa_1,\kappa_1,1+\alpha; \ldots;
  \kappa_6,\kappa_6,1+\alpha).
\end{equation}
As well, we choose an $\R^6$-vector $\mu>0$ with components $\mu_i$ , from which we define
\begin{equation}
   \label{eq:hatmuGowdy}
  \hat\mu=(\mu_1,\mu_1,\mu_1-(1+\alpha)+\kappa_1;\ldots;\mu_6,\mu_6,\mu_6-(1+\alpha)+\kappa_6).
\end{equation}
With these constructions, we formulate a singular initial value problem for the 
first-order system \Eqsref{eq:firstordersystemGowdy} -- \eqref{eq:firstordersystem2Gowdy}, seeking solutions  of the form
\[U=U_*+W\]
with $W\in X_{\delta,\hat\kappa+\hat\mu,\infty}$. 

Our construction of $\hat \kappa$ and $\hat \mu$ and their use in defining the function spaces $X_{\delta,\hat\kappa+\hat\mu,\infty}$ in which the remainder term lives are motivated by the following considerations.  We write  the exponent of the remainder term function space as  the sum $\hat\kappa+\hat\mu$ since $\hat\kappa$ represents the $t$-powers of the leading order term and hence the remainder is of ``higher order'' as required if $\hat\mu>0$. This, however, leaves the $1$-components of $\hat\kappa$ undetermined since  $U^i_{*,1}=0$. We make the particular choice of the components in \Eqref{eq:kappak}  for $\hat\kappa$ because, as we explain below in detail, one finds that $\hat\kappa$ agrees with the vector of eigenvalues $\Lambda$ (see \Eqref{eq:listoflambdas}) for our system. It follows then from the 
eigenvalue condition \Eqref{eq:eigenvaluecondition} of the well-posedness result  (Theorem \ref{th:smoothexistenceN}) that      $\hat\mu>0$. The particular form of the $1$-components of $\hat\mu$ in \Eqref{eq:hatmuGowdy} is thus a consequence of the block diagonal condition of Theorem \ref{th:smoothexistenceN} which, as we see below, requires that the $-1$-, $0$- and $1$-components of $\hat\kappa+\hat\mu$ are the same. 

\begin{proposition}[Existence of solutions of the singular initial value problem of the first-order evolution system]
  \label{prop:existenceevoleqs}
 Let the space $\mathcal{Q}$ consist of the following functions: 
  \begin{enumerate}[label=\textit{(\roman{*})}, ref=(\roman{*})] 
  \item \label{cond:kP} 
    A function $k\in C^\infty(T^1)$ such that $0<k(x)<3/4$ for all $x\in T^1$.
  \item Functions $\xi_0,\xi_1\in C^\infty(T^1)$ such that $\xi_0(x)>\max\{0,2k(x)-1\}$ and $\xi_1(x)>0$ for all $x\in T^1$.
  \item \label{cond:mu} An exponent vector $\mu$ such that
    \begin{center}
    \begin{tabular}{rcccl}
      $\max\{0,2k(x)-1\}$ & $<$ & $\mu_4(x)$ & $<$ & $\min\{1,\xi_0(x)\}$,\\
      $\max\{0,2k(x)-1\}$ & $<$ & $\mu_5(x)$ & $<$ & $\min\{\xi_0(x),2k(x),2(1-k(x))\}$,\\
       $\max\{0,2k(x)-1\}$ & $<$ & $\mu_1(x)$ & $<$ & $\min\{\mu_4(x),\mu_5(x)\}$,\\
      $0$ & $<$ & $\mu_6(x)$ & $<$ & $\min\{\mu_5(x),\mu_1(x)+1-2k(x)\}$,\\
      $\mu_1(x)$ & $=$ & $\mu_2(x)$ & $=$ & $\mu_3(x)$,
    \end{tabular}
  \end{center}
    for all $x\in T^1$.
  \item \label{cond:data} Functions $g_{00*}, g_{11*}, R_*, E_*, Q_*, Q_{**}\in C^\infty(T^1)$ such that
    $-g_{00_*}$, $g_{11_*}$, $R_*$, and $E_*$ are strictly positive.
  \item \label{cond:FN} Smooth asymptotic gauge source function data $F_{10}(x)$, $F_0(t,x,u)$ and $F_1(t,x,u)$ such that the corresponding
    function operators (defined below in Remark \ref{PropRemark3}) $W\mapsto F_0[W]$, $W\mapsto D F_0[W]$, and
    $W\mapsto \partial_x F_0[W]$ are rational $(\hat\kappa+\hat\mu,-1+\xi_0,\infty)$-operators and $W\mapsto \hat F_1[W]$, $W\mapsto D \hat F_1[W]$ and
    $W\mapsto \partial_x \hat F_1[W]$ are rational $(\hat\kappa+\hat\mu,\xi_1,\infty)$-operators.
  \end{enumerate}
  Moreover, choose
  \begin{equation}
    \label{eq:choosegammas}
    \gamma_0(x)=\frac 12\left(3+k(x)^2\right),\quad 
    \gamma_1(x)=\frac 14\left(1+k(x)^2\right)
  \end{equation}
  in \Eqref{eq:choiceconstraintmultiples}. If the constant $\alpha$ in \Eqsref{eq:firstordervariablesGowdy} and \eqref{eq:defUSGowdy} is sufficiently negative, then there exists a unique solution $U$ of the first-order system \Eqsref{eq:firstordersystemGowdy} -- \eqref{eq:firstordersystem2Gowdy} of the form
  \[U=U_*+W\quad\text{for some}\quad W\in X_{\tilde\delta,\hat\kappa+\hat\mu,\infty},\]
  for some $\tilde\delta>0$, where $\hat\kappa$, $\hat\mu$ and $U_*$ are given by \Eqsref{eq:defUSGowdy} -- \eqref{eq:hatmuGowdy}. Moreover, the remainder $W$ is differentiable in time and we have $DW\in
  X_{\tilde\delta,\hat\kappa+\hat\mu,\infty}$. 
\end{proposition}

Before proving this Proposition, we note the following:
\begin{remark}
\label{PropRemark1}
Comparing  $\mathcal{Q}$ and $\mathcal{P}$, we see that the hypothesis for Proposition \ref{prop:existenceevoleqs}
is significantly more general than that of \Theoremref{th:maintheorem}. This is not surprising, since this proposition is concerned only with obtaining solutions to the evolution equations, while \Theoremref{th:maintheorem}
is concerned with solutions of the \textit{full} Einstein equations: the constraints as well as the evolution equations. 
\end{remark}
\begin{remark}
\label{PropRemark2}
While \Theoremref{th:maintheorem} simply asserts the existence of an exponent vector $\mu$ for which the results hold, in this proposition \Conditionref{cond:mu} provides estimates for $\mu$. One finds that these estimates are the source of the restriction  $0<k<3/4$ appearing in \Conditionref{cond:kP} here, as well in the hypothesis  of \Theoremref{th:maintheorem} (see Remark \ref{kcondition}). 
In particular, it is the second inequality in \Conditionref{cond:mu} which implies the necessary restriction $2k-1<2(1-k)$; i.e., $k<3/4$. 

We emphasize that both the upper and the lower bounds in the inequalities for $\mu$ in \Conditionref{cond:mu} are meaningful. On the one hand, the strongest uniqueness statement is obtained by choosing the components of $\mu$ as small as possible, thus  giving rise to the ``biggest'' space for the remainder quantities. 
On the other hand, a large choice of $\mu$ close to the upper bound yields the most precise description of the actual behavior of the remainder quantities at $t=0$.

As seen below, some of these upper bounds are not fully optimal yet. We note that the order of the inequalities in \Conditionref{cond:mu} corresponds to the order in which components of $\mu$ can be picked which satisfy the inequalities. 
\end{remark}

\begin{remark}
\label{PropRemark3}
Here, we define the function operators appearing in \Conditionref{cond:FN} (note Remark \ref{nometricdependence} above).  
Given the function operator $W\mapsto F_0[W]$ (the same holds for $W\mapsto F_1[W]$), we define the map
\[W\mapsto D F_0[W],\quad W\mapsto DF_0[W]\] 
by specifying the function $DF_0[W](t,x)$ for any sufficiently regular function $W$ as follows: (i) We apply the $D$-derivative to the function $F_0[W](t,x)$, and (ii) we replace $DW_{-1}$ everywhere by $W_{0}+\alpha W_{-1}$ in agreement with the definition of the first-order variables; see \Eqref{eq:firstordervariablesGowdy}. Since gauge source functions are only allowed to depend the coordinates and on the metric, but in particular not on its derivatives, the map $W\mapsto D F_0[W]$ constructed like this is indeed a function operator. This would not be the case if there were terms including $DW_0$ or $DW_1$ after taking the $D$-derivative.
The map $W\mapsto \partial_x F_0[W]$ is defined in the same way and for the same reason is a function operator.
\end{remark}
\begin{remark}
\label{PropRemark4}
In the polarized case ($Q_*=Q_{**}=0$) the inequalities for $\mu_6$ and thereby the non-trivial lower bounds for $\mu_4$, $\mu_5$ and $\mu_1$ disappear. Moreover, the condition $\mu_5<2(1-k)$ vanishes. As a consequence the asymptotic velocity $k$ is allowed to be an arbitrary real function. As well, there is no restriction $\xi_0>2k-1$. In the half-polarized case ($Q_*=0$), the restriction $\mu_5<2(1-k)$ disappears and hence $k$ is allowed to be any positive function. The lower bound for $\xi_0$ however remains.
\end{remark}

\emph{Proof of Proposition \ref{prop:existenceevoleqs}:} We begin by rewriting the first-order evolution system \Eqsref{eq:firstordersystemGowdy} -- \eqref{eq:firstordersystem2Gowdy} in a form which is consistent with the criteria for establishing well-posedness in  \Theoremref{th:smoothexistenceN}. To this end, we replace the matrix $\tilde{\mathbf n} $ in \Eqref{eq:tildenGowdy} by its leading-order  expression
\[  \tilde{\mathbf n}=\begin{pmatrix}
    -\alpha & -1 & 0 \\
    -(1-\alpha)\alpha & -1+\alpha &0\\
    0 & 0 & (1+\alpha) \frac{g_{00*}}{g_{11*}}
  \end{pmatrix}
\]
and we absorb the higher-order terms into the source term operator $\tilde f[U]$, whose components now become 
\[\tilde f[U]^i=
    \left(0, \frac{2t^2 \Xi }{U^2_{-1}} H^i+2\alpha \frac{U^3_{-1}U^i_1}{U^2_{-1}},-(1+\alpha) B[W] U^i_1\right)^T,\]
where 
\[W\mapsto B[W]=\frac{g_{00*} t^{(k^2-1)/2}+W^1_{-1}}{g_{11*} t^{(k^2-1)/2} +W^2_{-1}}-\frac{g_{00*}}{g_{11*}}.\]
We then define the reduced
source term operator in \Eqref{eq:reducedeq} as
\begin{equation}
  \label{eq:reducedsource1}
  W\mapsto {\mathscr {\tilde F}}(U_*)[W]=\tilde f[U_*+W]- \LPDE{U_*+W}{U_*}
\end{equation}
where 
\[\LPDE{U_*+W}{U_*}=S^0[U_*+W] DU_*+S^1[U_*+W]  t\partial_{x} U_*+\tilde N U_*\]
and where the matrix $\tilde N$ is determined by the new matrix $\tilde{\mathbf n}$ via \Eqref{eq:block2fullGowdy}. 
We obtain
\begin{align*}
  \LPDE{U_*+W}{U_*}=\Bigl(&0,\frac 14(3-4k^2+k^4)t^{(k^2-1)/2}g_{00*},0;
  0,\frac 14(3-4k^2+k^4)t^{(k^2-1)/2}g_{11*},0;\\
  &0,0,0;
  0,0,0;
  0,E_* k (1+k)t^{-k},0;
  0,2k(2k-1)Q_{**} t^{2k},0\Bigr)^T\\
  &+S^1[U_*+W]  t\partial_{x} U_*,
\end{align*}
where
\[(S^1[U_*+W]  t\partial_{x} U_*)^i=\left(0,-2t \frac{U^3_{-1}}{U^2_{-1}} \partial_x U^i_{0*},t \frac{U^1_{-1}}{U^2_{-1}} \partial_x U^i_{0*}\right)^T\]
for each $i=1,\ldots,6$. 

The idea, now (following the discussion in \Sectionref{sec:ReviewOfSIVP}), is to establish that a modified version of the reduced source term operator ${\mathscr {\tilde F}} $ defined above in \Eqref{eq:reducedsource1}, which we label ${\mathscr {F}} $, has suitable regularity properties, and then show that it follows from \Theoremref{th:smoothexistenceN} that  the equation
\begin{equation}
\LPDE{U_* +W}{W}= \Fred{U_*}{W}
\end{equation}
has unique solutions. We establish the existence of this modified reduced source term operator in the following lemma:

\begin{lemma}
    \label{lem:evol}
    Let $U_*$ be given by \Eqref{eq:defUSGowdy}, let $\kappa$, $\hat\kappa$, $\mu$, and $\hat\mu$ be given by \Eqsref{eq:kappak} -- \eqref{eq:hatmuGowdy}, and suppose in addition that \Conditionsref{cond:k} -- \ref{cond:FN} of \Propref{prop:existenceevoleqs} hold. Choose the $\R^{18 \times 18}$ matrix 
\[N=\diag(N_{01},N_{22},N_{33},N_{44}),\]
where
\begin{center}
\resizebox{\linewidth}{!}{%
$N_{01}=\begin{pmatrix}
  -\alpha  & -1 & 0 & 0 & 0 & 0 & 0 & 0 & 0 \\
 \frac{1}{2} a
   d_1 & d_2 & 0 &
   \frac{a b g_{00*}}{4 g_{11*}} &
   \frac{bg_{00*} }{2 g_{11*}} & 0 & 0 & 0 & -\frac{b g_{00*}}{g_{11*}} \\
 0 & 0 & \frac{(\alpha +1) g_{00*}}{g_{11*}} & 0 & 0 & 0 & 0 & 0 &
   0 \\
 0 & 0 & 0 & -\alpha  & -1 & 0 & 0 & 0 & 0 \\
 0 & 0 & 0 & \frac{1}{4} a^2 &
   -k^2+\alpha +1 & 0 & 0 & 0 & -2 \\
 0 & 0 & 0 & 0 & 0 & \frac{(\alpha +1) g_{00*}}{g_{11*}} & 0 & 0 &
   0 \\
 0 & 0 & 0 & 0 & 0 & 0 & -\alpha  & -1 & 0 \\
 0 & 0 & e_1 & 0
   & 0 & -\frac{e_2 g_{00*}}{g_{11*}} & \frac{1}{2} a c_1 & c_2 & 0 \\
 0 & 0 & 0 & 0 & 0 & 0 & 0 & 0 & \frac{(\alpha +1)
   g_{00*}}{g_{11*}} 
\end{pmatrix}
,$
}
\end{center}
with
\begin{gather*}
a=-k^2+2 \alpha +1, \quad b=k^2-2 \gamma_0-1,\\ c_1=-k^2+\alpha +2 \gamma_1,\quad c_2=-\frac{3}{2} k^2+\alpha +2 \gamma_1+\frac{1}{2},\\
d_1=-k^2+\alpha +\gamma_0-1,\quad d_2=-\frac{3}{2}
   k^2+\alpha +\gamma_0-\frac{1}{2},\\
e_1=\frac{1}{4} \left(k^2-4 \gamma_1-5\right),\quad e_2=-\frac 14 \left(k^2-4 \gamma_1-1\right),
\end{gather*}
and
\[N_{22}=
\begin{pmatrix}
  -\alpha  & -1 & 0 \\
 (\alpha -1)^2 & \alpha -2 & 0 \\
 0 & 0 & \frac{(\alpha +1) g_{00*}}{g_{11*}}
\end{pmatrix},
\]
\[N_{33}=
\begin{pmatrix}
  -\alpha  & -1 & 0 \\
 (\alpha +k)^2 & \alpha +2 k & 0 \\
 0 & 0 & \frac{(\alpha +1) g_{00*}}{g_{11*}}
\end{pmatrix},
\]
\[N_{44}=
\begin{pmatrix}
 -\alpha  & -1 & 0 \\
 \alpha  (\alpha -2 k) & \alpha -2 k & 0 \\
 0 & 0 & \frac{(\alpha +1) g_{00*}}{g_{11*}}
\end{pmatrix}.
\]
Then there exists an exponent vector $\hat\nu>\hat\kappa+\hat\mu$ and a rational $(\hat\kappa+\hat\mu,\hat\nu,\infty)$-operator $W\mapsto \Fred{U_*}{W}$ such that
\[W\mapsto -\tilde N\cdot W+{\mathscr {\tilde F}}(U_*)[W]=-N\cdot W+\Fred{U_*}{W}.\]
  \end{lemma}
The proof of this lemma--in particular, the claimed regularity of the operator $W\mapsto \Fred{U_*}{W}$--follows directly from a computer-aided algebraic computation of this operator from the above definition. The details of the computer algebra code used are given in \Appendixref{sec:computeralgebra}. It is important to note that this proof is fully rigorous; numerical approximations do not play a role. 
We also note that while we could state explicit estimates for the exponent $\hat\nu$, such estimates are not needed to complete the proof of Proposition \ref{prop:existenceevoleqs}, which we complete here.

Using the results of this lemma, we rewrite  the first-order system \Eqsref{eq:firstordersystemGowdy} -- \eqref{eq:firstordersystem2Gowdy} in the Fuchsian form (see \Eqref{eq:pde})
\begin{equation}
  \label{eq:hasl}
  S^0 DW+S^1 t\partial_{x} W+N W=\Fred{U_*}{W}.
\end{equation}
We need to verify that this system  is indeed a smooth quasilinear symmetric hyperbolic Fuchsian system according to \Defref{def:quasilinearlimitN}. It is clear that $U_*\in C^\infty((0,\delta]\times T^n)\cap X_{\delta,\hat\kappa,\infty}$ and that all objects in the equations depend smoothly on their arguments on the relevant domains. Moreover, all function operators are rational. 
The matrix $S^0_0(U_*)$ can be constructed from $S^0$ in \Eqref{eq:block2fullGowdy} by replacing $-U^1_{-1}/U^2_{-1}$ in \Eqref{eq:s0Gowdy} by $-g_{00*}/g_{11*}$. It follows from \Conditionref{cond:data} of \Propref{prop:existenceevoleqs} that  this matrix is symmetric (in fact diagonal) and positive definite. Using the techniques in \Appendixref{sec:computeralgebra}, it is then straightforward to show that $W\mapsto S^0(W)-S^0_0(U_*)$ is a $(\hat\kappa+\hat\mu,\zeta,\infty)$-operator for some $\zeta>0$ (see \Eqref{S01} of \Defref{def:quasilinearlimitN}). In the same way, we can show that $W\mapsto t S^1(W)$ is a $(\hat\kappa+\hat\mu,\zeta,\infty)$-operator (see \Eqref{Sa} of \Defref{def:quasilinearlimitN}). This together with \Lemref{lem:evol} establishes that our evolution system \Eqref{eq:hasl} is indeed a smooth quasilinear symmetric hyperbolic rational-Fuchsian system.

We are now ready to apply \Theoremref{th:smoothexistenceN} and hence prove \Propref{prop:existenceevoleqs}. From the above constructions of the exponent vector $\hat\mu$ and of the matrices $S^0$, $S^1$ and $N$ in \Lemref{lem:evol}, it is clear that our system is block diagonal with respect to $\hat\mu$ and it is clear that $\hat\mu$ is ordered. The ordered vector of eigenvalues $\Lambda$ of the matrix $\NODE$ (see \Eqsref{eq:defNODE} and \eqref{eq:listoflambdas}) is found to be
  \begin{equation}
  \label{eq:LambdaGowdyPre}
  \begin{split}
    \Lambda=\Bigl(&(1-k^2)/2,\gamma_0-1-k^2,-1-\alpha; (1-k^2)/2,(1-k^2)/2,-1-\alpha;\\
&(1-k^2)/2,2\gamma_1-k^2,-1-\alpha;-1,-1,-1-\alpha;k,k,-1-\alpha;\\
&-2k,-2k,-1-\alpha\Bigr).
\end{split}
\end{equation}
If we therefore choose $\gamma_0$ and $\gamma_1$ as in \Eqref{eq:choosegammas}, it follows that
\begin{equation}
  \label{eq:LambdaGowdy}
  \begin{split}
    \Lambda=\Bigl(&(1-k^2)/2,(1-k^2)/2,-1-\alpha; (1-k^2)/2,(1-k^2)/2,-1-\alpha;\\
&(1-k^2)/2,(1-k^2)/2,-1-\alpha;-1,-1,-1-\alpha;k,k,-1-\alpha;\\
&-2k,-2k,-1-\alpha\Bigr).
\end{split}
\end{equation}
The condition \Eqref{eq:eigenvaluecondition} in \Theoremref{th:smoothexistenceN} is therefore satisfied for \textit{every} exponent vector $\hat\mu>0$ so long as $\alpha$ has been chosen sufficiently negative. 
This completes the proof of \Propref{prop:existenceevoleqs}.

\textbf{Carrying out Step 1c:}
Suppose that $U=U_*+W$ is any smooth solution of the first-order system \Eqsref{eq:firstordersystemGowdy} -- \eqref{eq:firstordersystem2Gowdy} with $U_*$ given by \Eqsref{eq:vecuSmetric}--\eqref{eq:defUSGowdy} with smooth data, and with $W,DW\in X_{\tilde\delta,\hat\kappa+\hat\mu,\infty}$ for $\hat\kappa$ given by \Eqsref{eq:kappak}--\eqref{eq:defKappaGowdy} and for any $\hat\mu>0$. We assume here that $k(x)\in (0,1)$ for all $x\in T^1$. We can then define
  \begin{equation*}
    u:=(U^1_{-1},\ldots,U^6_{-1})^T,\quad u_*:=(U^1_{-1*},\ldots,U^6_{-1*})^T,\quad 
    w:=(W^1_{-1},\ldots,W^6_{-1})^T.
  \end{equation*}
  Clearly, we have $u=u_*+w$ with $w, Dw\in X_{\tilde\delta,\kappa+\mu,\infty}$; in fact, $u\in C^\infty((0,\tilde\delta]\times T^1)$. In \Appendixref{sec:2ndorder21storder} we argue that this vector $u$ is a solution of the original second-order system \Eqref{eq:waveblubb} if and only if the six quantities
\begin{equation}
  \label{eq:constraintGowdy}
    C_1^i:=U_1^i-t\partial_{x}U^i_{-1}
  \end{equation}
vanish identically (see \Eqref{eq:constraint}). We also argue  in  \Appendixref{sec:2ndorder21storder} that since $U$ satisfies the first-order system \Eqsref{eq:firstordersystemGowdy} -- \eqref{eq:firstordersystem2Gowdy},  
these quantities must satisfy the subsidiary system 
\begin{equation}
  \label{eq:constraintpropGowdy}
  D C^i_1-(1+\alpha) C^i_1=0;
\end{equation}
 cf.\ \Eqref{eq:constraintprop}. This subsidiary system yields six decoupled linear homogeneous Fuchsian ordinary differential equations for which we can formulate a suitable singular initial value problem and then apply \Theoremref{th:smoothexistenceN} to this problem. \Theoremref{th:smoothexistenceN} and the homogeneity of \Eqsref{eq:constraintpropGowdy} imply that for each choice of the index $i$,  the unique solution of \Eqref{eq:constraintpropGowdy} contained in the space $X_{\delta,1+\alpha,\infty}$ is $C^i_1\equiv 0$. The quantities $C_1^i$ given by the vector $U$ by \Eqref{eq:constraintGowdy} are elements of the  space $X_{\tilde\delta,\kappa_i+\mu_i,\infty}$ (recall that $\mu_i<1$ is a consequence of \Conditionref{cond:mu} in \Propref{prop:existenceevoleqs}). If we therefore choose
the free constant $\alpha$ to be sufficiently negative (recall that this is consistent with the hypothesis of \Propref{prop:existenceevoleqs}), we can achieve that $1+\alpha < \kappa_i+\mu_i$ for all $i$. The unique solution of \Eqref{eq:constraintpropGowdy} in $X_{\tilde\delta,\kappa_i+\mu_i,\infty}$ is therefore indeed $C^i_1\equiv 0$.
We have thus derived the following statement.
\begin{proposition}
  \label{prop:existenceevoleqs2ndorder}
 Let $U=U_*+W$ be any solution of the first-order system \Eqsref{eq:firstordersystemGowdy} -- \eqref{eq:firstordersystem2Gowdy} with $U_*$ given by \Eqsref{eq:vecuSmetric}--\eqref{eq:defUSGowdy} by smooth data, and with $W,DW\in X_{\tilde\delta,\hat\kappa+\hat\mu,\infty}$ for $\hat\kappa$ given by \Eqsref{eq:kappak}--\eqref{eq:defKappaGowdy} and for any $\hat\mu>0$. We further assume that $k(x)\in (0,1)$ for all $x\in T^1$.
Then
  \begin{equation*}
    u:=(U^1_{-1},\ldots,U^6_{-1})^T
  \end{equation*}
  is a solution in $C^\infty((0,\tilde\delta]\times T^1)$ of the second-order system \Eqref{eq:waveblubb} of the form $u=u_*+w$ with
  \begin{equation*}
    u_*:=(U^1_{-1*},\ldots,U^6_{-1*})^T,\quad 
    w:=(W^1_{-1},\ldots,W^6_{-1})^T
  \end{equation*}
  where $w, Dw\in X_{\tilde \delta,\kappa+\mu,\infty}$. Moreover, for each $i=1,\ldots,d$, we have
  \[U^i_{1}=W^i_{1}=t\partial_x u^i,\]
  and hence
  \begin{equation}
    \label{eq:nexttermspatial}
    W^i_{1}=t\partial_x u^i_*+\tilde W^i_{1},
  \end{equation}
  with $\tilde W^i_{1}\in X_{\tilde\delta,\kappa_i+1+\tilde\mu_i,\infty}$ for some $\tilde\mu_i>0$.
\end{proposition}

It is evident that \Propref{prop:existenceevoleqs2ndorder} in particular applies to all solutions $U$ of \Propref{prop:existenceevoleqs}. Note, however, that some of the assumptions needed for \Propref{prop:existenceevoleqs}--for example the restriction $k\in (0,3/4)$--are not necessary here. 

\subsection{Verifying that the spacetimes are solutions}
\label{SptmsAreSolns}

In Section \ref{constructionof}, we have used the singular initial value problem to show that for a choice of asymptotic data contained in $\mathcal{P}$, one can develop a spacetime which satisfies the system \Eqref{eq:wgEinstEqsN} and matches this choice of asymptotic data. There is no guarantee, however, that this spacetime is a vacuum solution of  the Einstein equations. In this subsection, as the second major part of the proof of \Theoremref{th:maintheorem}, we show that indeed this spacetime is a vacuum solution. As noted above in the outline of the proof, before carrying through this verification that we have a proof, it useful to establish certain estimates for the shift, $g_{01}=U^3_{-1}$.

\textbf{Carrying out Step 2a:} 
We state and establish the desired estimates for the shift in the following result:
\begin{proposition}
  \label{prop:firstimprovement}   
  Suppose that in addition to the hypothesis of \Propref{prop:existenceevoleqs}, the function $F_{10}$ in \Conditionref{cond:FN} satisfies
\begin{equation}
  \label{eq:F10}
  F_{10}=\frac{1}{2} \left(
-2 k\frac{ E_*'}{E_*}
+4 k E_*^2 Q_{**} Q_{*}'
+\left(1-k^2\right)\frac{R_*'}{R_*}
-3\frac{g_{00*}'}{g_{00*}}
+\frac{g_{11*}'}{g_{11*}}
\right).
\end{equation}
Then the solution $U$ whose existence is asserted by \Propref{prop:existenceevoleqs} has the property that there exists an exponent scalar $\gamma>0$ 
such that the shift quantities
$U^3_{-1},U^3_{0},U^3_{1}$ are contained in $X_{\tilde\delta,(k^2-1)/2+1+\gamma,\infty}$.
\end{proposition}

The proof of this proposition proceeds as follows. Presuming that the hypothesis 
of \Propref{prop:existenceevoleqs} and \Eqref{eq:F10} hold, we denote the solution of the first-order system asserted by \Propref{prop:existenceevoleqs} by $\hat U$.
It follows from \Eqref{eq:nexttermspatial} that we can write $\hat U$ as $\hat U_*+\hat W$ with $\hat U_*$ taking the form
\begin{equation}
  \label{eq:bl1}
  \hat U_*=(\hat U_*^1,\ldots,\hat U_*^6)^T,\quad \hat U_*^i=(u^i_*, Du^i_*-\alpha u^i_*,t\partial_x u^i_*)^T,
\end{equation}
where $u_*$ is given by \Eqref{eq:vecuSmetric}. It further follows that $\hat W$ and $D\hat W$ are contained in $X_{\tilde\delta,\hat\kappa+\hat\mu,\infty}$ with 
\begin{equation}
\label{eq:DefHatKappa}
  \hat\kappa:=(\kappa_1,\kappa_1,\kappa_1; \ldots;
\kappa_6,\kappa_6,\kappa_6)
\end{equation}
where  $\kappa_1$,\ldots, $\kappa_6$ are given by \Eqref{eq:kappak},
and with 
\begin{equation}
  \label{eq:bl3}
  \begin{split}
  \hat\mu=(&\hat\mu_1,\hat\mu_1,\hat\mu_1+1;
\hat\mu_2,\hat\mu_2,\hat\mu_2+1;
\hat\mu_3,\hat\mu_3,\hat\mu_3+1;\\
&
\hat\mu_4,\hat\mu_4,\hat\mu_4+1;\hat\mu_5,\hat\mu_5,\hat\mu_5+1;
\hat\mu_6,\hat\mu_6,\hat\mu_6+1),
\end{split}
\end{equation}
where the quantities $\hat\mu_i$ are presumed to satisfy the inequalities in \Conditionref{cond:mu} in \Propref{prop:existenceevoleqs} with $\mu_i$ replaced by $\hat\mu_i$.

The basic idea is now to solve \Eqsref{eq:firstordersystemGowdy} -- \eqref{eq:firstordersystem2Gowdy} with the same data as in \Propref{prop:existenceevoleqs}, but now \textit{only for the shift quantities $U^3_{-1}$, $U^3_{0}$, $U^3_{1}$}, and to incorporate the PDEs for these quantities into 
 a singular initial value problem with improved exponents.  In doing this, we note that 
the less than optimal exponent for the shift quantities in \Propref{prop:existenceevoleqs} is a consequence of the restrictive block diagonal condition which is needed for the complete system. If instead we presume that all components of $U$ are known---i.e., if we set $U=\hat U$ except for the components $U^3_{-1}$, $U^3_{0}$ and $U^3_{1}$ and  if we then throw away all of the evolution equations from \Eqsref{eq:firstordersystemGowdy} -- \eqref{eq:firstordersystem2Gowdy} except for  the ones for $U^3_{-1}$, $U^3_{0}$ and $U^3_{1}$---then  the block diagonal condition becomes less restrictive, as we see below.  This reduced system of PDEs can be  rewritten as a first-order evolution system for the  ``unknowns'' $U^3_{-1}$, $U^3_{0}$, and $U^3_{1}$ only, with all of the matrices and coefficients  determined by the other components of $\hat U$:
\[\mathbf s^0 D
\begin{pmatrix}
  U^3_{-1}\\
  U^3_{0}\\
  U^3_{1}
\end{pmatrix}
+\mathbf s^1 t\partial_x
\begin{pmatrix}
  U^3_{-1}\\
  U^3_{0}\\
  U^3_{1}
\end{pmatrix}
+
\tilde{\mathbf n}\begin{pmatrix}
  U^3_{-1}\\
  U^3_{0}\\
  U^3_{1}
\end{pmatrix}
=\tilde g[U^3_{-1}, U^3_{0}, U^3_{1}].
\]
Here $\tilde g$ is some source term  (which we note is quite lengthy).
We now consider the singular initial value problem for these equations for $U^3_{-1}$, $U^3_{0}$, and $U^3_{1}$ with vanishing leading order terms
\[U^3_{-1}=W^3_{-1}\in X_{\tilde\delta,\kappa_3+\mu_3,\infty},\quad
U^3_{0}=W^3_{0}\in X_{\tilde\delta,\kappa_3+\mu_3,\infty},\quad
U^3_{1}=W^3_{1}\in X_{\tilde\delta,\kappa_3+\mu_3,\infty},\]
where $\mu_3>0$ is thus far unspecified.
Clearly, $U^3_{-1}=\hat U^3_{-1}$, $U^3_{0}=\hat U^3_{0}$, $U^3_{1}=\hat U^3_{1}$ is a solution of this singular initial value problem if $\mu_3\le\hat\mu_3$.
Using only the available information concerning  the components of $\hat U$ which is implied by \Eqsref{eq:bl1} -- \eqref{eq:bl3} together with \Eqref{eq:F10}, we can show (as a consequence of \Theoremref{th:smoothexistenceN}) that this singular initial value problem has a unique solution, provided that
\[0<\mu_3<1+\min\{\hat\mu_1,\hat\mu_6,\xi_1\}.\] 
We notice that if we were to not assume \Eqref{eq:F10},
then we would find the same statement (as above) for 
\[0<\mu_3<1.\]
We may, as a first step, choose $\mu_3\le \hat\mu_3$ (which is always smaller than $1$), and then use uniqueness to conclude that the particular choice $U^3_{-1}=\hat U^3_{-1}$, $U^3_{0}=\hat U^3_{0}$, $U^3_{1}=\hat U^3_{1}$ 
is the only solution of this 
singular initial value problem, as expected. 
We may then choose $\mu_3$ to be 
a bit larger than one, which implies that this solution indeed has the property asserted by \Propref{prop:firstimprovement}.

\textbf{Carrying out Step 2b:} 
To verify that the spacetimes constructed above in Section \ref{constructionof} are solutions of the Einstein equations, it is sufficient to show that, in terms of a chosen coordinate system, the quantities $\mathcal D_i$  vanish on these spacetimes. We show this here by setting up a singular initial value problem for $\mathcal D_i$ with vanishing leading-order data. 

We start by using the definition \Eqref{eq:DiDef} together with expressions \Eqsref{eq:blockdiagmetric} -- \eqref{eq:g33} for the Gowdy symmetric metric and expressions \Eqref{eq:asympwavegaugemainthm} for the gauge source functions to obtain the following formulas for the quantities $\mathcal D_i$:
\begin{equation}
  \label{eq:expressionDiGowdy}
  \begin{split}
  \mathcal D_0&=-\frac 1t+F_0+\frac{R_t}{R}+\frac{g_{01} g_{00,x}
    -g_{00} g_{01,x}
    +\frac{1}{2} g_{00} g_{11,t}
    -\frac{1}{2} g_{11} g_{00,t}}
  {g_{00} g_{11}-g_{01}^2}\\
 \mathcal D_1&=F_1+F_{10}+\frac{R_x}{R}
+\frac{g_{01} g_{11,t}
-g_{11} g_{01,t}
+\frac{1}{2} g_{11} g_{00,x}
-\frac{1}{2} g_{00} g_{11,x}}{g_{00} g_{11}-g_{01}^2}\\
\mathcal D_2&=\mathcal D_3=0.
\end{split}
\end{equation}
If we then differentiate these formulas with respect to $t$, replacing second time derivatives of $g_{ij}$ by means of the Einstein evolution equations \Eqref{eq:wavepdeNGowdy} and \Eqref{eq:wavepdeNGowdy2}, we obtain corresponding (lengthy) formulas for $D\mathcal D_0$ and $D\mathcal D_1$. Based on these formulas, we now verify that the leading order terms in the quantities  $\mathcal D_0$, $\mathcal D_1$, $D\mathcal D_0$, and $D\mathcal D_1$ (which we refer to collectively as the ``gauge-violation quantities'') all vanish, so long as the asymptotic data satisfy a certain asymptotic constraint condition, \Eqref{eq:constraintpre1}.\footnote{We note that this asymptotic constraint condition is included in the hypothesis of our main result, \Theoremref{th:maintheorem}.}  
\begin{lemma}
  \label{lem:Dis}
  Suppose that in addition to the hypothesis of \Propref{prop:existenceevoleqs}, the function $F_{10}$ in \Conditionref{cond:FN} of \Propref{prop:existenceevoleqs} satisfies \Eqref{eq:F10}.
  Let $U$ be the solution of the first-order evolution equations asserted by \Propref{prop:existenceevoleqs}.
Then, there exists an exponent scalar $\gamma>0$ such that the corresponding constraint violation quantities satisfy
\begin{equation}
\label{eq:Difunctsp1}
\mathcal D_0, D\mathcal D_0 \in X_{\tilde\delta,-1+\gamma,\infty}.
\end{equation}
If in addition to the above conditions, the asymptotic data satisfies 
\begin{equation}
  \label{eq:constraintpre1}
  F_{10}=-\frac{g_{00*}'}{2g_{00*}}+\frac{g_{11*}'}{2g_{11*}}-\frac{R_*'}{R_*},
\end{equation}
then
\begin{equation}
  \label{eq:Difunctsp2}
\mathcal D_1, D\mathcal D_1\in X_{\tilde\delta,\gamma,\infty}.
\end{equation}
\end{lemma}

Before proving this lemma, we note the following:
\begin{remark}
It follows immediately from \Eqref{eq:Difunctsp1} and \Eqref{eq:Difunctsp2} that the leading order terms of the constraint-violation quantities vanish. We stress that we obtain these conclusions \emph{only} if the asymptotic data satisfy both \Eqref{eq:F10} and \Eqref{eq:constraintpre1}. In particular, 
if \Eqref{eq:F10} holds but \Eqref{eq:constraintpre1} is violated, we can show that $\mathcal D_1\in X_{\tilde\delta,0,\infty}$. Hence \Eqref{eq:constraintpre1} can be interpreted as the condition which makes $\mathcal D_1$ vanish in leading order at $t=0$. 

We can write the two asymptotic constraints \Eqsref{eq:F10} and \Eqref{eq:constraintpre1} in the following form:
\begin{align*}
  \label{eq:asymptoticconstraints}
   \frac{g_{00*}'}{g_{00*}}&=-k\frac{ E_*'}{E_*}+2 k E_*^2 Q_{**} Q_{*}'+\frac{3-k^2}{2}\frac{R_*'}{R_*},\\
   \frac{g_{11*}'}{g_{11*}}&=\frac{g_{00*}'}{g_{00*}}+2 \frac{R_*'}{R_*}+2F_{10}.
\end{align*}

The first two of these equations is the origin of the integral constraint \Eqref{eq:integralconstraintN} for the asymptotic data $g_{00*}, E_*, Q_*, Q_{**}$ in \Theoremref{th:maintheorem} and for \Eqsref{eq:RSconst}. The second equation
is equivalent to \Eqref{eq:gaugeconstraint}.
We remark that if one uses the more common parametrization of the asymptotic data $E_*=e^{P_{**}}$ and $g_{00*}=-e^{\Lambda_{**}/2}$, if one imposes the ``conformal gauge condition'' $g_{00*}=-g_{11*}$ (which is usually part of the areal gauge assumption) and and if one sets $F_{10}=0$, then these conditions imply 
\begin{equation}
  \label{eq:GowdyArealConstraint}
  R_*'=0,\quad \Lambda_{**}'=-2k(P_{**}'-2 e^{2P_{**}}Q_{**} Q_{*}').
\end{equation}
These formulas are familiar for the singular initial value problem of Gowdy solutions in areal gauge \cite{Rendall:2000ki,Beyer:2010foa}.
\end{remark}

\emph{Proof of Lemma  \ref{lem:Dis}:} We presume that the hypothesis of \Propref{prop:existenceevoleqs} and \Eqref{eq:F10} both hold. As a consequence of \Propref{prop:firstimprovement}  and \Eqref{eq:nexttermspatial}, we can argue (as in the proof of \Propref{prop:firstimprovement}) that $U$ can be written as $U_*+W$, with $U_*$ given by 
\begin{equation}
  \label{eq:bll1}
  U_*=(U_*^1,\ldots,U_*^6)^T,\quad U_*^i=(u^i_*, Du^i_*-\alpha u^i_*,t\partial_x u^i_*)^T,
\end{equation}
where $u_*$ is given by \Eqref{eq:vecuSmetric}. Moreover, it follows that $W$ and $DW$ are contained in $X_{\tilde\delta,\hat\kappa+\hat\mu,\infty}$, with 
\begin{equation}
  \label{eq:bll2}
  \begin{split}
  \hat\kappa:=(&\kappa_1,\kappa_1,\kappa_1;
  \kappa_1,\kappa_1,\kappa_1;
  \kappa_2,\kappa_2,\kappa_2;
  \kappa_3+1,\kappa_3+1,\kappa_3+1;\\
  &\kappa_4,\kappa_4,\kappa_4;
  \kappa_5,\kappa_5,\kappa_5;
\kappa_6,\kappa_6,\kappa_6),
\end{split}
\end{equation}
with $\kappa$ given by \Eqref{eq:kappak},
and with 
\begin{equation}
  \label{eq:bll3}
  \begin{split}
  \hat\mu=(&\mu_1,\mu_1,\mu_1+1;
\mu_2,\mu_2,\mu_2+1;
\mu_3,\mu_3,\mu_3+1;\\
&
\mu_4,\mu_4,\mu_4+1;\mu_5,\mu_5,\mu_5+1;
\mu_6,\mu_6,\mu_6+1).
\end{split}
\end{equation}
All of the quantities $\mu_i$ except for $\mu_3$ are assumed to satisfy the inequalities in \Conditionref{cond:mu} in \Propref{prop:existenceevoleqs}, while $\mu_3$ is  some (sufficiently small) positive exponent (see \Propref{prop:firstimprovement}). Using techniques similar to those we have applied above to derive expansions of operator functions, we verify (i) that $\mathcal D_0\in X_{\tilde\delta,-1+\gamma,\infty}$ with 
\[\gamma=\min\{\xi_0,\mu_1\},\]
and (ii) that
\[\mathcal D_1-\left(F_{10}+\frac{g_{00*}'}{2g_{00*}}-\frac{g_{11*}'}{2g_{11*}}+\frac{R_*'}{R_*}\right)\]
is contained  in $X_{\tilde\delta,\gamma,\infty}$ for 
\[\gamma=\min\{\xi_1,\mu_1,\mu_3\}.\]
Similar arguments apply to the more complicated expressions of $D \mathcal D_0$ and $D\mathcal D_1$, thereby completing the proof of \Lemref{lem:Dis}.

Having now derived the function spaces \Eqsref{eq:Difunctsp1} and \eqref{eq:Difunctsp2} for the constraint violation quantities (presuming that the asymptotic constraints hold), our next step is to show that the constraint violation quantities must be \textit{identically zero}. We know that the constraint violation quantities associated with a solution of the evolution equations must satisfy the constraint propagation system \Eqref{eq:ConstraintPropagationEquationN} with \Eqsref{eq:choiceconstraintmultiples} and \eqref{eq:choosegammas}. This system takes the form \Eqref{eq:wavepdeN}; i.e,
\begin{equation}
  \label{eq:wavepdeNGowdyConstr}
      \sum_{k,l=0}^1g^{kl}\partial_{x^k}\partial_{x^l} \mathcal D_i= 2 H_{i} 
\end{equation}
where $H_i$ is determined by \Eqref{eq:ConstraintPropagationEquationN}.
We wish to replace this second-order PDE system with a first-order system (so that we can apply our results concerning the well-posedness of singular initial value problems); we do this using the ideas discussed in Appendix A. More specifically, we combine $\mathcal D_0$ and $\mathcal D_1)$ into a vector 
\[v=(\mathcal D_0,\mathcal D_1)^T,\]
we label the first derivatives of components of $v$ in the form
\begin{equation}
  \label{eq:firstordervariablesGowdyConstr}
  V_{-1}^i:=v^i,\quad V_0^i:=Dv^i-\alpha v^i,\quad V_1^i:=t\partial_{x}
  v^i,
  \quad V^i:=(V^i_{-1},V_0^i,V_1^i)^T,
\end{equation}
for $i=1,2$,
where $\alpha$ is a constant to be fixed below (possibly different from the constant $\alpha$ discussed above), and we combine these to form the six-dimensional vector
\begin{equation}
  \label{eq:firstordervariables2GowdyConstr}
  V:=(V^1,V^2)^T.
\end{equation}
One readily verifies that 
the second-order system for $v$  implies a first-order system for $V$ of the form \Eqsref{eq:firstordersystem}--\eqref{eq:firstordersystem2}; i.e.,
\begin{equation} 
  \label{eq:firstordersystemGowdyConstr}
  S^0(t,x) DV(t,x)+S^1(t,x) t\partial_{x} V(t,x)+N(t,x) V(t,x)=0,
\end{equation}
where 
\begin{equation*}
  S^0=\diag (\mathbf s^0,\mathbf s^0),\quad
  S^1=\diag (\mathbf s^1,\mathbf s^1),
\end{equation*}
with $\mathbf s^0$ and $\mathbf s^1$  given by \Eqsref{eq:s0Gowdy} and \eqref{eq:s1Gowdy}.
The special form of the third term in \Eqref{eq:firstordersystemGowdyConstr}
is a consequence of  linear homogeneity. 

To show that the singular initial value problem for $V$ based on \Eqref{eq:firstordersystemGowdyConstr} is well-posed, we need to verify a certain fall-off rate for the matrix $N$. To do this, we first note that it follows from its construction (based on  \Eqref{eq:wavepdeNGowdyConstr}) that $N$ is fully determined by the components of the first-order vector $U$ corresponding to the given solution of the evolution equations.
More specifically, presuming that
the hypothesis of \Propref{prop:existenceevoleqs} and \Eqref{eq:F10} both  hold, we argue (as in the proof of 
\Lemref{lem:Dis}) that $U$ is of the form $U_*+W$ with $U_*$ given \Eqref{eq:bll1} and $u_*$ given by \Eqref{eq:vecuSmetric}. Moreover, $W$ and $DW$ are in $X_{\tilde\delta,\hat\kappa+\hat\mu,\infty}$ with $\hat\kappa$ given by \Eqref{eq:bll2} and $\kappa$ by \Eqref{eq:kappak},
and with $\hat\mu$ given by \Eqref{eq:bll3}
where all quantities $\mu_i$, except for $\mu_3$, are assumed to satisfy the inequalities stated in  \Conditionref{cond:mu} of \Propref{prop:existenceevoleqs}, while $\mu_3$ is some (sufficiently small) positive exponent. It follows that
\[N-N_0\in X_{\tilde\delta,\zeta,\infty}\]
for some $\zeta>0$ where
\begin{equation}
  \label{eq:N0Constr}
  N_0:=
\begin{pmatrix}
  -\alpha  & -1 & 0 & 0 & 0 & 0 \\
  (\alpha +1)^2 & \alpha +2 & 0 & 0 & 0 & \frac{g_{00*}}{g_{11*}} \\
  0 & 0 & (1+\alpha)\frac{g_{00*}}{g_{11*}} & 0 & 0 & 0 \\
  0 & 0 & 0 & -\alpha  & -1 & 0 \\
  0 & 0 & -2 & \alpha  (\alpha +1) & \alpha +1 & 0 \\
  0 & 0 & 0 & 0 & 0 & (1+\alpha)\frac{g_{00*} }{g_{11*}} 
\end{pmatrix};
\end{equation}
cf.~\Eqref{Ncond} of \Defref{def:quasilinearlimitN}.
Noting that the eigenvalues of the matrix $(S^0_0)^{-1} N_0$ are 
\begin{equation}
  \label{eq:eigenvalconstr1}
  \Lambda=(1,1,-1-\alpha;0,1,-1-\alpha),
\end{equation}
we determine that it follows from  \Theoremref{th:smoothexistenceN}  that \emph{if} we can show that the vector field $V$ satisfies the regularity condition
\begin{equation}
  \label{eq:VSpace}
  V\in X_{\tilde\delta,(\gamma,\gamma,\gamma,\gamma,\gamma,\gamma),\infty}
\end{equation}
then the singular initial value problem for $V$ based on \Eqref{eq:firstordersystemGowdyConstr}
has a unique solution for any $\gamma>0$;
here the particular structure of the exponent in \Eqref{eq:VSpace} is a consequence of the block diagonal condition. Since $V\equiv 0$ solves this singular initial value problem, it  follows (presuming \Eqref{eq:VSpace})
that this is the only solution of \Eqref{eq:firstordersystemGowdyConstr} in the space \Eqref{eq:VSpace}. 

As noted in Section \ref{sec:outlinemainproof} (following the preview of Step 1c),  in fact there is a mismatch between the regularity for $V$ provided by Lemma \ref{lem:Dis}-- as stated explicitly in \Eqref{eq:Difunctsp1} and \Eqref{eq:Difunctsp2}--and that needed for the singular initial value problem to be well-posed, as stated in 
\Eqref{eq:VSpace}. To compare these, we note that the regularity provided by Lemma \ref{lem:Dis} can be stated as
\begin{equation}
  \label{eq:VSpace2}
  V\in X_{\tilde\delta,(-1+\gamma,-1+\gamma,-1+\gamma,\gamma,\gamma,\gamma),\infty}.
\end{equation}

To show that in fact the conditions hypothesized in Lemma \ref{lem:Dis} \emph{are} sufficient to guarantee the regularity \Eqref{eq:VSpace}, we use arguments very similar to those used in Proposition \ref{prop:firstimprovement}  (in Step 2a) to prove the required enhanced regularity of the shift. Specifically, presuming that 
the hypothesis of \Lemref{lem:Dis} and \Eqref{eq:constraintpre1} hold, we readily determine that \Eqref{eq:Difunctsp2} implies estimates for $\mathcal D_1$ and  $D\mathcal D_1$ which are sufficient for \Eqref{eq:VSpace}. The required estimates for $\mathcal D_0$ and  $D\mathcal D_0$ are not so immediate. To obtain them, we 
choose  \textit{any} function $\mathcal D_0$ which is consistent with the above stated regularity  (we do not, however, choose  $\mathcal D_0\equiv 0$ since this is one of the things we are aiming to show) and we work with \Eqref{eq:firstordersystemGowdyConstr} as an evolution system \textit{only} for $V^0=(V^0_{-1},V^0_{0},V^0_{1})^T$; i.e., we delete the evolution equations for the now \textit{given} quantity $V^1=(V^1_{-1},V^1_{0},V^1_{1})^T$, but keep the evolution equations for the now \textit{unknown} quantity
$V^0=(V^0_{-1},V^0_{0},V^0_{1})^T$.

For this smaller system here with a hence less restrictive block diagonal condition we are led to conclude that this singular initial value problem has a unique solution
\[V^0=(V^0_{-1}, V^0_{0},V^0_{1})\in X_{\tilde\delta,(-1+\eta,-1+\eta,-1+\eta),\infty}\]
provided $0<\eta<1+\gamma$. 
In analogy with the arguments in the proof of \Propref{prop:firstimprovement},
one yields the sought improved estimates. 
We thus have verified that indeed \Eqref{eq:VSpace} holds.  The argument leading to the vanishing of $\mathcal D_i$ follows, and we have the following result:

\begin{proposition}
  The constraint violation quantities in \Lemref{lem:Dis} vanish identically, i.e.,
\[\mathcal D_0\equiv \mathcal D_1\equiv 0\]
on the whole existence interval $(0,\tilde\delta]$ of the solution $U$.
\end{proposition}

As noted above, the vanishing of the constraint violation quantities implies that the spacetimes built in Steps 1a-1c are solutions of the vacuum Einstein equations.

\subsection{Verifying that the spacetimes exhibit AVTD behavior}
\label{sec:AVTDGowdy}

To complete the proof of our main result, Theorem \ref{th:maintheorem}, it remains to show that these Gowdy spacetimes exhibit AVTD behavior in terms of the general (wave-type) coordinates employed in the constructions described in Steps 1a-1c. We do this here. 

\textbf{Carrying out Step 3:}
The concept of solutions of Einstein's equations exhibiting AVTD behavior has been formalized in \cite{Isenberg:1990gn,Isenberg:1999ba,Isenberg:2002ku} through  the introduction of a ``velocity term dominated'' (VTD) PDE system. The VTD system consists of both evolution and constraint equations and is constructed, with respect to a given system of coordinates, by dropping the spatial derivative terms in the Einstein evolution equations and in the Hamiltonian constraint. A solution of the full Einstein equations is said to be AVTD with respect to the chosen system of coordinates if it approaches, in a suitable norm, a solution to the VTD system (or its leading order).

We recall the usual procedure in the literature for establishing the existence of AVTD solutions. The VTD evolution system forms a spatially parameterized system of ODEs. It may be possible to find explicit solutions to this system, although knowledge of the leading order behavior is sufficient to establish the VTD property. One establishes the existence of solutions with AVTD behavior by first setting up a singular initial value problem for the evolution equations, where the leading order term is chosen to be in agreement with the VTD solution. In a subsequent step, one formulates a singular initial value problem for the Hamiltonian and Momentum constraint violation quantities. It follows that provided certain constraints on the spatially-varying asymptotic data hold, one obtains unique solutions to the full Einstein system. Moreover, it follows that if the singular initial value problem takes the Fuchsian form \Defref{def:quasilinearlimitN}, then by definition these solutions must be AVTD.

To facilitate this  discussion, it is useful to introduce a bit of terminology concerning systems \Eqref{eq:pde}. By the corresponding \keyword{truncated system} we mean the first-order system formed from \Eqref{eq:pde} by dropping the spatial derivative terms $\sum_{a=1}^n \Ssna(U) t \partial_a U$. The following corollary 
of \Theoremref{th:smoothexistenceN} concerns existence of solutions to the singular initial value problem for such a truncated system.

\begin{corollary}[of \Theoremref{th:smoothexistenceN}]
\label{cor:ExistenceUniquenessForTruncatedSystem}
Suppose that for a system \Eqref{eq:pde} the conditions of \Theoremref{th:smoothexistenceN} have been met for some leading order term $U_*(t,x)$, with asymptotic data (parametrized by a set of quantities $\delta$ and $\mu$) satisfying certain constraints $\mathcal C$. Then the corresponding truncated system also satisfies the conditions of \Theoremref{th:smoothexistenceN} with the same leading order term $U_*(t,x)$, and with the same ($\mu$ and $\delta$ parametrized) asymptotic data satisfying $\mathcal C$. Thus there exists a family of solutions with leading order term $U_*(t,x)$, parametrized by the same set of asymptotic data, to the corresponding truncated system.
\end{corollary}

This corollary follows from the definition of a Fuchsian system, \Defref{def:quasilinearlimitN}. For such a Fuchsian system the function operators $W \mapsto t \Ssna(W)$, which are the coefficients of the spatial derivative terms, are (by definition) $(\mu,\zeta,\infty)$-operators for some exponent vector  $\zeta>0$. As such, the spatial derivative terms are guaranteed to be higher order in $t$ than the terms which match the decay of $W$, and thus these terms do not constrain the singular decay rate of the solutions obtained in \Theoremref{th:smoothexistenceN}. The singular initial value problem for the truncated system can be seen as just a special case of \Theoremref{th:smoothexistenceN}, with $\zeta$ approaching  infinity. 

As we discuss now, the AVTD property of the solutions under consideration in \Theoremref{th:maintheorem} is almost a consequence of \Corref{cor:ExistenceUniquenessForTruncatedSystem}. To see this, we consider the family of solutions constructed as discussed in \Theoremref{th:maintheorem}, with functions $F_0$ and $F_1$ (cf. \Eqref{eq:asympwavegaugemainthm}) in function spaces parametrized by $\xi_0, \xi_1$ and satisfying \Conditionref{cond:xi}. 
For any such choice of gauge, this family (which we label as $\mathcal S_{\xi, F}$)  is parametrized by the set of asymptotic data, $\mathcal P$, satisfying the relations \Conditionsref{cond:k} - \ref{cond:dataTh} of that theorem. In particular, these solutions satisfy the evolution equations \Eqsref{eq:firstordersystemGowdy} -- \eqref{eq:firstordersystem2Gowdy}, and the hypotheses of \Propref{prop:existenceevoleqs}. An application of \Corref{cor:ExistenceUniquenessForTruncatedSystem} verifies the existence of a corresponding family of solutions, which we denote by $\widetilde {\mathcal S}_{\xi, F}$, to the corresponding truncated system with the same functions $F_0$ and $F_1$ in function spaces parametrized by $\xi_0, \xi_1$ and which is parametrized by the same set of asymptotic data $\mathcal P$. This argument shows that each of the solutions to the full Einstein system obtained in \Theoremref{th:maintheorem} approaches a corresponding solution of the first-order truncated evolution equations.

We now argue that the first-order truncated system is almost equivalent to the VTD system associated to the Einstein equations. One might worry that the system has been truncated at first-order, not second-order, and hence the spatial derivatives are still there in the form of the first-order fields $U^i_1$. One finds that in the truncated system the equations for the $U^i_1$ decouples from the other equations and forms a homogeneous system of ODE. It follows from the  uniqueness of the solutions in \Corref{cor:ExistenceUniquenessForTruncatedSystem} that $U^i_1 = 0$ is the only solution, and as a consequence this system is equivalent to the first-order system formed from the VTD equations. 

In our application there is an additional subtlety due to the (non-standard) definition of $Q$ in \Eqsref{eq:g23} and \eqref{eq:g33}. As a result of this definition, the truncated system corresponding to  \Eqsref{eq:firstordersystemGowdy} -- \eqref{eq:firstordersystem2Gowdy} (with, in addition, $U^i_1 = 0$ in accord with the argument above) differs from the first-order VTD system by terms proportional to $Q_*'(x)$ and $Q_*''(x)$. This simply reflects that fact that in our choice of the variable $Q$ we have already ``accounted for'' part of the VTD leading order term, and moreover, it is straightforward to check that the truncated system with these terms removed has the same existence properties as the full truncated system.

In summary we have established that for any fixed set of asymptotic data and gauge source functions consistent with the constraints and restrictions of \Theoremref{th:maintheorem} the two singular initial value problems, (i) for the full Einstein equations (asserted by \Theoremref{th:maintheorem}), and, (ii) for the VTD equations, each have a solution. Because both solutions have the same asymptotic data and their remainders are controlled by the same $t$-dependent norms, their difference approaches zero in the sense of the function spaces in \Theoremref{th:maintheorem}. We have therefore established that the solutions given in \Theoremref{th:maintheorem} are AVTD.

Does this demonstration that the solutions $\mathcal S_{\xi, F}$ exhibit AVTD behavior include the Einstein constraint equations as well as the evolution equations? In fact it does; this follows from \Eqref{eq:EinsteinConstraintsAndConstraintViolationQuantities}, which relates the Einstein constraints to the vanishing of the generalized wave gauge constraint violation quantities. It follows from this relation that the vanishing of $\partial_t \mathcal D_0$ and $\partial_t \mathcal D_1$ to leading order is  equivalent to the constraints vanishing at leading order. For areal coordinates, this equivalence is manifest in \Eqref{eq:GowdyArealConstraint}.

\section{Main solution space and relationship between coordinate systems}
\label{sec:coordinaterelationship}

It is well-established that in terms of areal coordinates, $T^3$-Gowdy solutions generically exhibit AVTD behavior. Since the main result of this work is the demonstration that there are Gowdy solutions which exhibit AVTD behavior in terms of generalized wave coordinates as well, it is useful to examine the relationship between AVTD behavior as seen in alternative coordinate systems, and how such features change under coordinate transformations from one system to another. We do this analysis here; for brevity, we omit some of the technical details.

We recall that it follows from \Theoremref{th:maintheorem} that for each choice of data $k$, $g_{11*}$, $g_{00**}$, $R_{*}, E_*, Q_*, Q_{**}$, and for each choice of the gauge source functions of the form \Eqref{eq:asympwavegauge} which are consistent with the restrictions of the theorem, there is a unique metric $g$ which solves Einstein's vacuum equations and which is given in the unique coordinate representation \Eqsref{eq:leadinorderterms} -- \eqref{eq:leadinordertermsII}.
For the present discussion, we consider any two such metrics $g^{(1)}$ and $g^{(2)}$ of \Theoremref{th:maintheorem} to be  the same --- and hence we write $g^{(1)}=g^{(2)}$ --- if and only if they are determined by the same data and the same gauge source functions. We consider two sets of data and gauge source functions as the same if and only if they are the same in the sense of functions, respectively. We stress that in the  discussion here we are intentionally \textit{not} considering  diffeomorphism-equivalence classes of solutions of \Theoremref{th:maintheorem}.  

\newcommand{\Ar}{\text{\textsc{\tiny A}}}
\newcommand{\SolSet}{\mathcal S}
Let $\SolSet$ be the set of all solutions obtained from the theorem in this sense.
Let $\SolSet^\Ar\subset\SolSet$ be the subset of \keyword{areal solutions}; i.e., the subset of $\SolSet$ which is determined by the special data $g_{00**}=1$, $R_*=1$, $F_{10}=0$ and $F_0\equiv F_1\equiv 0$, and where all other data functions are subject to the standard areal Gowdy constraint
\[\int_0^{2\pi}\left(
        -k(x)\frac{ E_*'(x)}{E_*(x)}+2 k(x) E_*^2(x) Q_{**}(x) Q_{*}'(x)\right)dx=0.\]
The two constraints \Eqsref{eq:RSconst} and \eqref{eq:gaugeconstraint} then imply that
\begin{equation*}
  -g_{00*}(x)=g_{11*}(x)= e^{\int_0^{x}\bigl(
    -k(\xi)\frac{ E_*'(\xi)}{E_*(\xi)}+2 k(\xi) E_*^2(\xi) Q_{**}(\xi) Q_{*}'(\xi)\bigr)d\xi}.
\end{equation*} 
Comparing  \Theoremref{th:maintheorem} with areal-coordinate AVTD results \cite{Rendall:2000ki,Beyer:2010foa}, we  conclude that all elements in $\SolSet^\Ar$ have the property $R\equiv t$, $g_{01}\equiv 0$ and $g_{00}\equiv -g_{11}$, and hence these metrics are indeed represented in areal coordinates. 

In order to distinguish areal coordinates in the following discussion from any other coordinate system consistent with \Theoremref{th:maintheorem} we refer to the former as $(t^\Ar,x^\Ar,y^\Ar,z^\Ar)$. We shall demonstrate now that
 the following 
type of coordinate transformations plays an important role for \Theoremref{th:maintheorem}:
\begin{equation}
  \label{eq:coordtrafo}
  \begin{split}
    t^\Ar(t,x,y,z)&=t^\Ar(t,x)=(\tau(x)+f_0(t,x)) t,\\
    x^\Ar(t,x,y,z)&=x^\Ar(t,x)=x+h_0(x)+(h_1(x)+f_1(t,x))t^2,\\ 
    \quad y^\Ar(t,x,y,z)&=y,\quad z^\Ar(t,x,y,z) =z,
  \end{split}
\end{equation}
for so far unspecified smooth $2\pi$-periodic (with respect to $x$) functions $\tau(x)$, $h_0(x)$, $h_1(x)$, $f_0(t,x)$ and $f_1(t,x)$ which have the property that $\tau(x)>0$ and $h'_0(x)>-1$ for all $x\in T^1$, and that $f_0$ and $f_1$ are in $X_{\delta,\eta,\infty}\cap C^\infty((0,\delta]\times T^1)$ for some $\eta>0$. If $\delta>0$ is sufficiently small as we always assume, the map $(t,x,y,z)\mapsto (t^\Ar,x^\Ar,y^\Ar,z^\Ar)$ is invertible on $(0,\delta]\times T^3$ and hence indeed a coordinate transformation.
Each such  coordinate transformation maps any element $g^\Ar\in\SolSet^\Ar$ to some $g$; that is,  it transforms any metric from its representation in areal coordinates $(t^\Ar,x^\Ar,y^\Ar,z^\Ar)$ to its representation in some other coordinates $(t,x,y,z)$.
We can show under suitable further technical assumptions on $f_0$ and $f_1$ that 
\begin{align}
  \label{eq:metric1}
    g_{00}(t,x)&=-g_{00*}^\Ar(x+h_0(x))\tau^2(x)
    (\tau(x))^{(k^2(x+h_0(x))-1)/2}\cdot\\
    &\qquad\cdot{t}^{(k^2(x+h_0(x))-1)/2} (1+\ldots),\notag\\
    g_{11}(t,x)&=g_{00*}^\Ar(x+h_0(x))(1+h_0'(x))^2
    (\tau(x))^{(k^2(x+h_0(x))-1)/2}\cdot\\
    &\qquad\cdot{t}^{(k^2(x+h_0(x))-1)/2} (1+\ldots),\notag\\  
  \label{eq:metric01}
    g_{01}(t,x)&=g_{00*}^\Ar(x+h_0(x))
    \left(2h_1(x)(1+h_0'(x))-\tau(x)\tau'(x)\right)\cdot\\
    &\qquad\cdot(\tau(x))^{(k^2(x+h_0(x))-1)/2}
    {t}^{(k^2(x+h_0(x))+1)/2} (1+\ldots),\notag\\
  g_{02}&\equiv g_{03}\equiv g_{12}
  \equiv g_{13}\equiv 0,
\end{align}
and
\begin{align}
  R(t,x)&=t\, \tau(x) (1+\ldots),\\
  E(t,x)&=E_*^\Ar(x+h_0(x)) {\tau(x)}^{-k(x+h_0(x))}{t}^{-k(x+h_0(x))} (1+\ldots),\\
  \label{eq:metricQ}
  Q(t,x)&=Q_{**}^\Ar(x+h_0(x)) {\tau(x)}^{2k(x+h_0(x))}{t}^{2k(x+h_0(x))}(1+\ldots),
\end{align}
where the data which determine the original areal solution $g^\Ar$ are labelled with ${}^\Ar$.
Here we write 
\[H_1(t,x)=H_2(t,x)+\ldots\] 
for two arbitrary functions $H_1$ and $H_2$ provided $H_1-H_2$ is a function in $X_{\delta,\epsilon,\infty}\cap C^{\infty}((0,\delta]\times T^1)$ for some $\epsilon>0$. In order to make the following discussion fully rigorous we would need to give precise estimates of the higher-order terms represented by ``$\ldots$'' above in terms of $\eta$. It is not difficult to obtain those, but for brevity we do not discuss them here. One can show that if we choose $\eta$ ``sufficiently large'' then everything in the following is justified rigorously. 

Further calculations, similar to those which led to \Eqsref{eq:metric1} -- \eqref{eq:metricQ}, allow us to find
\begin{align}
    \Gamma_0&=-\frac1t+\ldots,\\
    \label{eq:Gamma1trafo}
    \Gamma_1&=-\frac{2 h_1 (1+h_0')}{\tau^2}-\frac{\tau'}{\tau}+\frac{h_0''}{1+h_0'}+\ldots.
\end{align}

Now, if $g^\Ar\in\SolSet^\Ar$ and hence the metric represented by $g^\Ar$ is a solution of the vacuum equation, the same is true for the image metric of the coordinate transformation above which we continue to refer to as $g$. Nevertheless, this $g$ is \textit{not} always in $\SolSet$. In particular we observe that if the leading term in \Eqref{eq:metric01} does not vanish, then \Eqsref{eq:leadinordertermsshift} and \eqref{eq:remainderstheorem} of our theorem are violated. 
However, $g$ must be a solution of \Theoremref{th:maintheorem} and hence be an element of $\SolSet$ if, (i), the asymptotic data for $g$ implied by the leading terms of  \Eqsref{eq:metric1} -- \eqref{eq:metricQ} satisfy the constraints \Eqsref{eq:integralconstraintN} -- \eqref{eq:gaugeconstraint} of \Theoremref{th:maintheorem}, and if, (ii), $\eta$ is sufficiently large so that \Conditionref{cond:xi} of \Theoremref{th:maintheorem} is met. This is a consequence of uniqueness. It turns out that this is the case
if and only if, (i), 
\begin{equation}
  \label{eq:blabblab}
  2h_1(x)(1+h_0'(x))-\tau(x)\tau'(x)=0,
\end{equation}
i.e., the leading term in \Eqref{eq:metric01} indeed vanishes, and (ii), $\eta$ is sufficiently large.

Next, let $\Xi$ denote the set of all coordinate transformations of the form above which is consistent with \eqref{eq:blabblab} and for which $\eta$ is sufficiently large. As we have seen, any element $\phi$ of $\Xi$ defines a map
\[\Phi_{(\phi)}:\SolSet^\Ar\rightarrow \SolSet_{(\phi)},\quad g^\Ar\mapsto g\]
given by \Eqsref{eq:metric1} -- \eqref{eq:metricQ},
where
\[ \SolSet_{(\phi)}:=\Phi_{(\phi)}(\SolSet^\Ar)\subset\SolSet.\]
We can show that for each $\phi\in\Xi$, this map $\Phi_{(\phi)}$ is bijective. It is obvious that $\SolSet_{(\phi)}$ is a proper subset of $\SolSet$ and that
\[\cup_{\phi\in\Xi} \SolSet_{(\phi)}\subset\SolSet.\]

An interesting question, which arises from this but which we shall not fully answer in this paper, is whether
\begin{equation}
  \label{eq:conj}
  \cup_{\phi\in\Xi} \SolSet_{(\phi)}=\SolSet.
\end{equation}
If the answer is no, then there exists at least one solution guaranteed by \Theoremref{th:maintheorem} which cannot be obtained from an areal solution by means of a coordinate transformation $\phi\in\Xi$. Does $\SolSet$ possibly contain solutions which are \textit{geometrically} distinct from areal solutions? Or, could the equality in \Eqref{eq:conj} fail just because the class of coordinate transformations $\Xi$ is not general enough? 

In order to approach such questions, we need to study whether it is possible to construct a coordinate transformation which maps an arbitrary solution $g$ in $\SolSet$ to an areal solution $g^\Ar$ in $\SolSet^\Ar$. Here we can exploit the fact that  in the generalized wave formalism this coordinate transformation map must be a solution of the following system of wave equations (cf.\ \Eqref{eq:coordwaveeq}):
\begin{align*}
  \Box_g {t^\Ar}(t,x)&=-{\mathcal F^\Ar}^0(t^\Ar(t,x),x^\Ar(t,x))=\frac1{t^\Ar(t,x)} {g^\Ar}^{00}(t^\Ar(t,x),x^\Ar(t,x))\\
  &=\frac1{t^\Ar(t,x)}\left(g^{00}(t,x)\left(\frac{\partial t^\Ar}{\partial t}\right)^2+2g^{01}(t,x) \frac{\partial t^\Ar}{\partial x}\frac{\partial t^\Ar}{\partial t}+g^{11}(t,x)\left(\frac{\partial t^\Ar}{\partial x}\right)^2\right),\\
\Box_g {x^\Ar}(t,x)&=0.
\end{align*}
The idea would be to formulate a singular initial value problem for this system with the leading-order behavior given by \Eqref{eq:coordtrafo} and \Eqref{eq:blabblab}. If this turned out to be successful and certain further technical details were met we would be able to decide whether \Eqref{eq:conj} is true.

Finally, another consequence of \Eqref{eq:coordtrafo} and \Eqref{eq:blabblab}. It suggests that the assumption that the shift $g_{01}$ be $o(t^{(k^2+1)/2})$ which we were forced to make in the course of the proof of our main theorem is possibly of purely technical nature. Namely, \Eqref{eq:metric01} shows that a metric with a shift which does not satisfy this assumption can easily be generated via the coordinate transformation \Eqref{eq:coordtrafo} simply by violating \Eqref{eq:blabblab}.


\section*{Acknowledgements}
This material is based upon work supported by the National Science Foundation under Grant No. OISE-1210144 while author EA was visiting the University of Otago on an EAPSI Fellowship in 2012. Portions of this paper were written during a visit of the author PLF at the University of Otago with the financial support of FB's ``Divisional assistance grant'' and during a visit of the author FB to the Universit\'e Pierre et Marie Curie with the support from the Agence Nationale de la Recherche via the grant 06-2--134423.  
This material is also based upon work supported by the National Science Foundation under Grant No.~0932078 000, while all authors were in residence at the Mathematical Science Research Institute in Berkeley, California, during the semester/year of 2013. The authors also gratefully acknowledge the support of the Institute Henri Poincar\'e during the 2015 Program of the Centre \'Emile Borel on Mathematical General Relativity.
The author FB was also partly funded by a University of Otago Research Grant 2013. The author PLF also gratefully acknowledges support from the Simons Center for Geometry and Physics, Stony Brook University, in January 2015. JI is partially supported by NSF grant PHY-1306441.

\addcontentsline{toc}{section}{References}


\addcontentsline{toc}{section}{Appendix}
\appendix

\section{First-order reduction of second-order wave equations}
\label{sec:2ndorder21storder}

In this portion of the appendix, we describe the reduction of certain quasilinear second-order PDE systems to first-order symmetric hyperbolic PDE systems. The second-order systems we consider here take the form 
\begin{equation}
  \label{eq:wavepdeN}
  \begin{split}
    \sum_{i,j=0}^ng^{ij}(t,x,&u(t,x),\partial_t u(t,x),\partial_{x^a} u(t,x)) \partial_{x^i}
    \partial_{x^j}u(t,x)\\
    &\qquad\qquad\qquad= 2H(t,x,u(t,x),\partial_t u(t,x),\partial_{x^a} u(t,x))
  \end{split}
\end{equation}
where $x^i$ (for $i$ running from $0$ to $n$, with $x^0=t$) are local coordinates on an $(n+1)$-dimensional manifold $M$, where $u(x^i)=u(t,x^a)=u(t,x)$ (for  $a=1,\ldots,n$) is an unknown $\R^d$-valued function on $M$, where $g^{ij}(t,x,u(t,x))$ are components of the inverse of a Lorentz-signature metric on $M$, and where $H(t,x,u(t,x),\partial_t u(t,x),\partial_{x^a} u(t,x))$ is an $\R^d$-valued function of the indicated variables. We presume that $g^{ij}$ and $H$ are specified function of the indicated quantities, and the system \Eqref{eq:wavepdeN} is to be solved for $u$.

Since we are in particular interested in systems with degeneracies at $t=0$, we find it useful to multiply both sides of \Eqref{eq:wavepdeN}  by $t^2$, and then rewrite \eqref{eq:wavepdeN} in the form
\begin{equation}
  \label{eq:wavepde2N}
  D^2u
  -2\sum_{a=1}^n{G^{0a}} t\partial_{x^a} Du
  -\sum_{a,b=1}^nG^{ab} t^2\partial_{x^a}\partial_{x^b}u
  - D u
  = \frac{2t^2}{g^{00}} H,
\end{equation}
where $D:=t\partial_t$, $G^{ij}:=-\frac{g^{ij}}{g^{00}}$, and $G:=(G^{ab})$.

To obtain first-order form, we define the variables
  \begin{equation}
    \label{eq:firstordervariables}
    \begin{split}
      U_{-1}^J&:=u^J,\quad U_0^J:=Du^J-\alpha u^J,\quad U_a^J:=t\partial_{x^a}
      u^J,\\
      \quad U^J&:=(U^J_{-1},U_0^J,U_1^J,\ldots,U^J_n)^T,
    \end{split}
  \end{equation}
  for $J=1,\ldots,d$ and $a=1,\ldots,n$, and we define the
  $(n+2)\cdot d$-vector
  \begin{equation}
    \label{eq:firstordervariables2}
    U:=(U^1,\ldots,U^d)^T.
  \end{equation}
  Here $\alpha$ is a constant, which is useful  in the Fuchsian analysis of these equations. 
  In terms of $U$, \Eqref{eq:wavepde2N} implies the first-order system
  \begin{equation}
    \label{eq:firstordersystem}
    S^0 DU+\sum_{a=1}^n S^a t\partial_{x^a} U+\tilde N U=\tilde f[U],
   \end{equation}
  with
  \begin{equation}
    \label{eq:block2full}
    S^0=\diag (\mathbf s^0,\ldots,\mathbf s^0),\quad
    S^a=\diag (\mathbf s^a,\ldots,\mathbf s^a),\quad
    \tilde N=\diag(\tilde{\mathbf n},\ldots,\tilde{\mathbf n}),
  \end{equation}
  where 
  \begin{equation}
    \label{eq:s0}
    \mathbf s^0=
    \begin{pmatrix}
      1 & 0 & 0\\
      0 & 1 & 0\\
      0 & 0 & G
    \end{pmatrix},
    \quad
    \mathbf s^a
    =
    \begin{pmatrix}
      0 & 0 & 0 & \ldots & 0\\
      0 & -2G^{0a} & -G^{1a} & \ldots & -G^{na}\\
      0 & -G^{1a} &  &  & \\
      \vdots & \vdots&  & \zeromatrix_n & \\
      0 & -G^{na} &  &  & 
    \end{pmatrix},
  \end{equation}
  and 
  \begin{equation}
    \label{eq:firstordern}
    \tilde{\mathbf n}=
    \begin{pmatrix}
      -\alpha & -1 & 0 & \ldots & 0\\
      -(1-\alpha)\alpha & -1+\alpha & 0 & \ldots &
      0\\
      0 & 0 & & & \\
      \vdots & \vdots& & -(1+\alpha) G & \\
      0 & 0 & & & \\
    \end{pmatrix},
  \end{equation}
 and with 
  \begin{equation}
    \label{eq:firstordersystem2}
    \begin{split}
    \tilde f[U]=
    \Biggl(&0, \frac{2t^2}{g^{00}} H^1+2\alpha \sum_{a=1}^nG^{0a} U^1_a,0,\ldots,0; \ldots; \\
      &0, \frac{2t^2}{g^{00}} H^d+2\alpha \sum_{a=1}^nG^{0a} U^d_a,0,\ldots,0\Biggr)^T.
  \end{split}
\end{equation}
This system is symmetric hyperbolic so long as the matrix $G$, which
 generally depends on the solution, is positive definite. We note that in these matrix equations, we use 
 $\zeromatrix_m$ to denote the $m\times m$-zero matrix.

It is relatively straightforward to verify the equivalence of the first-order system \Eqref{eq:firstordersystem}-\eqref{eq:firstordersystem2} and the original second-order system \Eqref{eq:wavepdeN}. In one direction, it follows from the derivation of system \eqref{eq:firstordersystem}-\eqref{eq:firstordersystem2} that if an $\R^d$-valued function $u$ is $C^2((0,\delta)\times T^3)$ and satisfies the second-order system \eqref{eq:wavepdeN}, then if we define $U$ by \Eqref{eq:firstordervariables}-\eqref{eq:firstordervariables2}, $U$ must satisfy the first-order system \Eqref{eq:firstordersystem}-\eqref{eq:firstordersystem2}. 

Going in the other direction, we consider an $\R^{(n+2)d}$ -valued function $U$ which is $C^1((0,\delta)\times T^3)$ and satisfies \Eqref{eq:firstordersystem}-\eqref{eq:firstordersystem2}. If we define  
 \begin{equation}
    \label{eq:first2second}
    u:=(U^1_{-1},\ldots,U^d_{-1})^T, 
  \end{equation}
we find that it is generally not a solution of \Eqref{eq:wavepde2N}. However, if all of the quantities $U^J_{-1}$ are $C^2$, and if the 
$n\cdot d$ functions
  \begin{equation}
    \label{eq:constraint}
    C_a^J:=U_a^J-t\partial_{x^a}U^J_{-1}
  \end{equation}
  vanish identically
  for all $a=1,\ldots,n$ and $i=1,\ldots,d$, then $u$ 
  is indeed a classical solution of the
  second-order system \Eqref{eq:wavepde2N}.
  
  In fact, it is sufficient that we know that the $C_a^J$ quantities vanish for a particular value $t_{*}\in (0,\delta)$. This follows immediately from the first-order ODE system  
  \begin{equation}
  \label{eq:constraintprop}
  D C^J_a-(1+\alpha) C^J_a=0
\end{equation}
for  $C_a^J$, which is implied by the first-order system  \Eqref{eq:firstordersystem}-\eqref{eq:firstordersystem2}. Clearly if  
 $C_a^J$ vanishes at $t=t_{*}$ and if satisfies the above linear homogenous ODE system, then it vanishes for all $t\in(0,\delta).$


\section{Some technical results and our computer algebra code}
\label{sec:computeralgebra}

Fix some $\delta>0$. As in Subsection \ref{classofeqns}, we consider $\R^d$-valued functions $u$,
 which can be written as $u_*+w$ for some fixed $u_*\in C^\infty((0,\delta]\times T^n)\cap X_{\delta,\kappa,\infty}$ and arbitrary functions $w\in X_{\delta,\kappa+\mu,\infty}$ for exponent $d$-vectors $\kappa$ and $\mu>0$. Let two function operators $w\mapsto f(w)$ and $w\mapsto g(w)$ be given. For the following it is useful to introduce
the notation $w\mapsto f(w)=g(w)+O(t^\nu)$ if the function operator $w\mapsto f(w)-g(w)$ is a $(\kappa+\mu,\nu,\infty)$-operator for some exponent $\nu$.

We consider the following algebraic operations involving function operators. The proofs of the following statements can be derived straightforwardly from the ideas in \cite{Ames:2012vz}.  
  \begin{description}
  \item [Sum of two function operators.] Let $\nu_1$ and $\nu_2$ be two exponent scalars. Let a scalar-valued $(\kappa+\mu,\nu_{1},\infty)$-operator $w\mapsto g_1(w)$ and a scalar-valued $(\kappa+\mu,\nu_{2},\infty)$-operator $w\mapsto g_2(w)$ be given. Then the map $w\mapsto g_1(w)+g_2(w)$  is a $(\kappa+\mu,\min\{\nu_{1},\nu_{2}\},\infty)$-operator\footnote{With a slight abuse of notation we write $\min\{\nu_{1},\nu_{2}\}$ for any smooth function $\nu(x)$ with has the property $\nu(x)<\min\{\nu_{1}(x),\nu_{2}(x)\}$ for every $x\in T^1$. Here we consider any two exponent scalars $\nu_1$ and $\nu_2$. Notice that we can always choose the difference between $\nu(x)$ and the actual (possibly non-differentiable) function $\min\{\nu_{1}(x),\nu_{2}(x)\}$ to be arbitrarily small.}. Moreover, for any two other function operators 
\[w\mapsto h_1(w):=g_1(w)+O(t^{\eta_1})\quad\text{and}\quad w\mapsto h_2(w):=g_2(w)+O(t^{\eta_2}),\]  
  for exponent scalars $\eta_1$, $\eta_2$,  we have
\[w\mapsto h_1(w)+h_2(w)=g_1(w)+g_2(w) +O(t^{\min\{\eta_1,\eta_2\}}).\]
\item [Product of two function operators.] Given the same function operators as before, the map $w\mapsto g_1(w)g_2(w)$ is a $(\kappa+\mu,\nu_{1}+\nu_{2},\infty)$-operator, and
\[w\mapsto h_1(w) h_2(w)=g_1(w) g_2(w)
+O(t^{\min\{\nu_1+\eta_2,\nu_2+\eta_1,\eta_1+\eta_2\}}).\]
\item [Inverse of a function operator.] Suppose that $w\mapsto P(w)$ is a scalar-valued $(\kappa+\mu,\zeta,\infty)$-operator for some $\zeta>0$. Then
$w\mapsto 1/(1+P(w))$ is a $(\kappa+\mu,0,\infty)$-operator, and
\[w\mapsto \frac{1}{1+P(w)}=1-P(w)+O(t^{2\zeta}).\]
Now let (i) $\eta, \gamma, \nu$ be exponent scalars with $\nu<\gamma<\eta$, (ii) $h_0$ be a function in $X_{\delta,\nu,\infty}$ such that $1/h_0\in X_{\delta,-\nu,\infty}$, and, (iii) 
$w\mapsto g(w)$ be a $(\kappa+\mu,\gamma,\infty)$-operator.
Suppose  
\begin{equation}
  \label{eq:denominatorform}
  w\mapsto P(w)=h_0+g(w)+O(t^\eta).
\end{equation}
Then, we have 
\begin{equation}
\label{eq:inversedenominatorform}
w\mapsto \frac{1}{P(w)}=\frac{1}{h_0} - \frac{g(w)}{h_0^2}+O(t^{-\nu+\min\{2(\gamma-\nu),\eta-\nu\}}).
\end{equation}
\end{description}

In our applications here,  all of the function operators are rational (see \Defref{def:rationalFOPs}) and hence are built using (possibly very many) terms each of which has a simple structure to which the algebraic rules above apply. Each term can be written as
\begin{equation}
  \label{eq:rationalfuncexp}
  w\mapsto \frac{H^{(1)}[w]}{H^{(2)}[w]},
\end{equation}
where both $w\mapsto H^{(1)}[w]$ and $w\mapsto H^{(2)}[w]$ are scalar polynomial function operators. More specifically, we can assume that 
there is a smooth function $P^{(1)}(t,x)$ in $X_{\delta,\nu,\infty}$ for some exponent scalar $\nu$ and non-negative integers $i_1$,\ldots $i_d$ such that
\begin{equation}
  \label{eq:polynFOPexp}
  H^{(1)}[w](t,x)=P^{(1)}(t,x)\cdot (u_{*1}(t,x)+w_1(t,x))^{i_1}\cdots (u_{*d}(t,x)+w_d(t,x))^{i_d}.
\end{equation}
This function operator can be analyzed by (i) considering the map $w\mapsto u_{*i}+w_i$ as a $(\kappa+\mu,\kappa_i+\mu_i,\infty)$-operator if $u_{*i}(t,x)=0$ for all $(t,x)\in (0,\delta]\times T^n$ or as a $(\kappa+\mu,\kappa_i,\infty)$-operator if $u_{*i}(t,x)\not=0$ for some $(t,x)\in (0,\delta]\times T^n$, and, (ii) applying the above algebraic rules. Similarly, we consider the ``trivial'' map $w\mapsto P^{(1)}$ as a $(\kappa+\mu,\nu,\infty)$-operator. 
For most of our applications, we want to prove that $H^{(1)}$ satisfies a ``linear expansion'' of the form
\begin{equation}
  \label{eq:polynFOPexp2} 
  H^{(1)}[w](t,x)=H^{(1)}_0(t,x)+\sum_{i=1}^d H^{(1)}_i(t,x) w_i(t,x)+O(t^\gamma)
\end{equation}
and we want to determine the functions $H^{(1)}_0(t,x)$, $H^{(1)}_1(t,x)$, \ldots, $H^{(1)}_d(t,x)$ explicitly and estimate the exponent scalar $\gamma$ in terms of $\kappa$ and $\mu$. In order to achieve this, we expand \Eqref{eq:polynFOPexp} using the algebraic rules above and ``linearize'' every product as follows
\[w\mapsto w_i w_j=O(t^{\kappa_i+\kappa_j+\mu_i+\mu_j}),\]
for each $i,j=1,\ldots,d$.
While this linearization is justified rigorously, it may not always give optimal results because in complicated expressions there may be important cancellations of nonlinear terms. In practice one may therefore end up with formally correct, but useless linear expansions.

Regarding the denominator $H^{(2)}[w]$ in \Eqref{eq:rationalfuncexp} we proceed in basically the same way as for the numerator. In general, $H^{(2)}[w]$ is a finite sum of terms of the form \Eqref{eq:polynFOPexp} and hence we can use the same algebraic rules and algorithm as above to derive an ``expansion'' of the form \Eqref{eq:polynFOPexp2}, i.e.,
\begin{equation}
  \label{eq:polynFOPexp3} 
  H^{(2)}[w](t,x)=H^{(2)}_0(t,x)+\sum_{i=1}^d H^{(2)}_i(t,x) w_i(t,x)+O(t^\gamma),
\end{equation}
for some possibly different exponent $\gamma$.
In doing this, the idea is to apply the above rule for the inverse of function operators and finally multiply the result with the numerator function operator \Eqref{eq:polynFOPexp2} using again the same rules. Eventually one obtains a ``linear expansion'' of the same form as in \Eqref{eq:polynFOPexp2}, but now for the \textit{full} function operator in \Eqref{eq:rationalfuncexp}
\begin{equation}
   \label{eq:rationalFOPexp} 
  \frac{H^{(1)}[w]}{H^{(2)}[w]}(t,x)=P_0(t,x)+\sum_{i=1}^d P_i(t,x) w_i(t,x)+O(t^\gamma),
\end{equation}
for an in  general again different exponent $\gamma$.

In practice, one needs to pay particular attention in applying the inverse rule above  because it only holds under strict assumptions.
Fortunately, we find that while there are very many different numerator function operators in our applications, only a few different denominator operators appear. Hence, we are able to check that the assumptions for the inverse rule hold explicitly for each of these operators.

\paragraph{Our computer algebra code.}
In practical applications we have to deal with function operators which consist of hundreds of terms of the form above. Each term can be processed by means of the simple algebraic rules discussed above. The analysis therefore becomes a very repetitive task which is performed very well by means of computer algebra. Indeed, we have implemented all the rules above and all the function operators which appear in our applications, using Mathematica. 
We stress that the results obtained in this way are fully rigorous and, in particular, no numerical approximation is used anywhere.


\end{document}